\documentclass[onecolumn,amsmath,amssymb,aps,pra]{revtex4-2}

\usepackage{graphicx}
\usepackage{dcolumn}
\usepackage{bm}
\usepackage{hyperref}
\usepackage{physics}
\usepackage{placeins} 
\begin{document}
	
	\title{Quantum Random Access Memory Implementation Using Photon-Photon Interaction in Rydberg Atomic Ensemble}
	
	\author{Avirup Chakraborty}
	\affiliation{S.N. Bose National Centre for Basic Sciences, \\JD Block, Sector-III, Bidhannagar, Kolkata, West Bengal-700106, India.}
	\email{avirup.chakraborty@bose.res.in}  
	
	\author{Shrabana Chakrabarti}
	\affiliation{Sister Nivedita University, \\DG 1/2, DG Block(Newtown), Action Area I, Newtown, Kolkata, Chakpachuria, West Bengal 700156, India.}
	\email{shrabana.cs@snuniv.ac.in}
	
	\date{\today}
	
	\begin{abstract}
		Quantum random access memory (qRAM) is crucial for overcoming data-loading bottlenecks in quantum machine learning; however, current physical implementations face severe scalability constraints. Traditional fanout designs demand exponential decoherence-prone gates, while bucket-brigade schemes require highly error-prone active switches. Motivated by these limitations, we propose a scalable qRAM architecture that fundamentally replaces active nodes with phase-encoded quantum walkers. Our methodology maps a discrete-time quantum walk onto a cavity quantum electrodynamics framework utilizing an electromagnetically induced transparency (EIT)-based Rydberg atomic ensemble. Inside hollow-core waveguides, strong Rydberg dipole-dipole interactions and a solenoidal magnetic field create a robust routing operator. This operator imparts precise, polarization-dependent phase shifts, steering circularly polarized probe pulses to target memory cells. Our results demonstrate that operating within a strong control field regime suppresses emergent spatial attenuation, ensuring cumulative transmission probabilities for highly scaled memory addresses. Ultimately, this parallelized architecture successfully optimizes spatial resources to static gates and temporal complexity to an optimal logarithmic scale of $\mathcal{O}(n\log(n+m))$ by requiring $\mathcal{O}(n+m)$ physical walkers, establishing a practical, fault-tolerant hardware pathway for advanced quantum computation implementations.
	\end{abstract}
	
	
	\maketitle
	
	\section{Introduction}
	
	Quantum computation originated from Turing machine proposals \cite{benioff1980}, many-body simulations \cite{feynman1982}, and formalized parallelism \cite{deutsch1985}, surpassing classical limits \cite{bernstein1993, berthiaume1992}. Algorithmic breakthroughs \cite{shor1994, grover1996} provide exponential linear algebra advantages, accelerating machine learning tasks like HHL clustering of $N$-dimensional vectors into $M$ states \cite{harrow2009, lloyd2013, aaronson2015}. To resolve HHL dataloading overheads \cite{aaronson2015, lloyd2013, biamonte2017}, Giovannetti proposed quantum random access memory (qRAM) \cite{giovannetti2008_1, giovannetti2008_2, giovannetti2008_3}:
	\begin{equation}
		\sum_j \psi_j |j\rangle_a \xrightarrow{\text{qRAM}} \sum_j \psi_j |j\rangle_a |D_j\rangle_d
	\end{equation}
	where $D_j$ is the $j$th cell content. However, physical implementations face scalability limits \cite{giovannetti2008_2, giovannetti2008_3, duan2003}: fanout routing along $2^n$ paths requires exponential decoherence-prone gates, while the bucket brigade requires $n = \log(N)$ steps across $\mathcal{O}(n)$ qutrits with error rates below $\mathcal{O}(n^{-2})$, complicating error correction.
	
	To overcome this, recent approaches leverage quantum random walks (QRWs) for coherent graph exploration \cite{aharonov1993, ambainis2001, aharonov2001, childs2009}. Asaka et al. proposed a discrete-time quantum walk (DTQW) qRAM on a binary tree. Routing particles via internal chirality states, $|0\rangle_c$ (left) and $|1\rangle_c$ (right), eliminates coherent interactions across $\mathcal{O}(n 2^n)$ nodes. They formulated address superposition retrieval as
	\begin{equation}
		\text{qRAM} : |0\rangle_A |0,0\rangle_B |0\rangle_C |0\rangle_D \mapsto \frac{1}{\sqrt{|A|}} \sum_{a \in A} |a\rangle_A |0,0\rangle_B |0\rangle_C |x(a)\rangle_D
	\end{equation}
	where $|0\rangle_A, |0\rangle_C$, and $|0\rangle_D$ denote the address, internal, and data state vectors, respectively, and $|w,l\rangle_B$ represents the walker's position within an $n$-level binary tree \cite{asaka2021, asaka2023_1, asaka2023_2}.
	
	Implementing unitary transformations without geometric phase gates bypasses solid-state cryogenic constraints \cite{schmidt2003, leibfried2003, kane1998, pashkin2003}. Photonic EIT systems provide superior isolation \cite{obrien2003, petrosyan2005, andre2002, gasparoni2004}, but lack strong two-photon interactions required for scalability \cite{agarwal2005, cote2006}. We resolve this using Rydberg phase gates \cite{jaksch2000}, where robust dipole-dipole interactions yield high-fidelity entanglement \cite{moller2008}. Unlike cross-phase modulation \cite{friedler2005}, our design couples Zeeman-split ground states to excited Rydberg states in a strong solenoidal magnetic field \cite{shahmoon2011, petrosyan2012, paredes2014, petrosyan2004}. Consequently, the resulting polaritons generate opposite polarization-dependent phases, steering quantum walkers towards the correct daughter nodes.
	
	We propose a QRW-based qRAM architecture that replaces decoherence-prone gates with quantum walkers, thereby minimizing unitary interactions. This parallelized framework reduces the operational complexity to $\mathcal{O}(n\log(n+m))$ for $n$ addresses and $m$ data qubits, utilizing $\mathcal{O}(n+m)$ resources. Physically, we introduce a cavity QED architecture that relies on photon-photon interactions. Phase-encoded walkers, driven by Rydberg dipole-dipole interactions, traverse a binary tree of phase-sensitive waveguides with cold atomic ensemble nodes exhibiting electromagnetically induced transparency. Eliminating exponential gates, this Rydberg blockade regime ensures superior scalability for numerous quantum computational applications \cite{farhi1998}. Ultimately, this study realizes three core unitary operations: routing, querying, and retrieving quantum walkers.
	
	\section{Theoretical Methodology}
	
	\subsection{Polarization-Dependent Phase Generation for Oppositely Circular Polarized Probes}
\begin{figure}[htbp]
		\centering
		\includegraphics[width=0.5\linewidth]{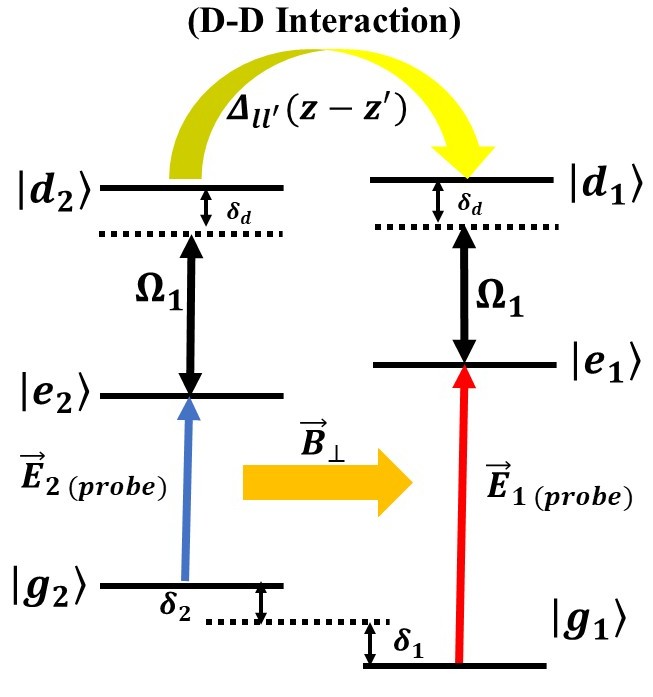}
		\caption{Double-ladder Electromagnetically Induced Transparency (EIT) scheme utilized within the Rydberg atomic ensemble.}
		\label{fig:fig1}
	\end{figure}

	To generate entanglement between two polarized probe components, we model cold atoms (Doppler broadening being less than the spin-relaxation rate) inside a hollow-core photonic crystal waveguide. A magnetic field $B$ Zeeman-splits the initially populated ground states $\left|g_1\right\rangle$ and $\left|g_2\right\rangle$ by $\hbar \Delta=g_F M_F \mu_B B$ ( $\mu_B$ : Bohr magneton, $g_F$ : gyromagnetic factor, $M_F$ : magnetic quantum number). The system employs a double-ladder EIT scheme (See Fig.~\ref{fig:fig1}) driven by counterpropagating weak probes and a strong control field. The atomic density is $\rho=\left(\pi w_a^2\right)^{-1} e^{-r_1^2 / w_a^2}(N / L)$. Ground states couple to intermediate states $\left|e_1\right\rangle$ and $\left|e_2\right\rangle$ via probe fields $\hat{\varepsilon}_{1,2}=$ $\sqrt{\hbar \omega_l / 2 \epsilon_0 V} e^{-r_1^2 / 2 \omega_l^2} \hat{E}_{1,2}$ with detuning $\delta_{1,2}=\omega_p-\omega_{e_{1,2} g_{1,2}}-k_p v \mp \Delta$. A strong control field (Rabi frequency $\Omega$ ) couples $\left|e_1\right\rangle$ and $\left|e_2\right\rangle$ to Rydberg states $\left|d_1\right\rangle$ and $\left|d_2\right\rangle$, detuned by $\delta_d=\omega_d-\omega_{e_{1,2} d_{1,2}}-k_d v+\Delta^{\prime}$, where $\Delta^{\prime}=$ $-\mu_B M_F^{\prime} g_F^{\prime} B$. Here, $w_f, w_a$ are Gaussian field/ensemble widths; $\omega_p, \omega_d$ are laser frequencies; $\omega_{e_{1,2} g_{1,2}}$ are resonant frequencies; $k_p, k_d$ are wavevectors; and $v$ is beam velocity. Rydberg states $\left|d_{1,2}\right\rangle$ experience a dipole-dipole shift $\Delta_{l l}\left(z-z^{\prime}\right)=C_{12} \frac{\left(1-3 \cos ^2 \theta\right)}{\left|z-z^{\prime}\right|^3}$, with $C_{12}=\frac{P_{d_1} P_{d_2}}{4 \pi \epsilon_0 \hbar} \quad\left(P_{d_{1,2}}\right.$ being dipole moments).We define $g=$ $\varrho_{\mathrm{eg}} \sqrt{\omega_p /\left(2 \hbar \epsilon_0 A L\right)}$ being the atom-field coupling strength, $\varrho_{\mathrm{eg}}$ being Clebsch-Gordan coefficient, $A$ being the cross-sectional area of the laser field and $L$ being the medium length. The system's Hamiltonian via density matrices is
	
	$$
	\begin{array}{r}
		H_{i n t}=\frac{\hbar N e^{-z_1^2 / w_n^2}}{L} \int d z\left[\delta_1 \hat{\sigma}_{g_1 g_1}+\delta_2 \hat{\sigma}_{g_2 g_2}+\int d^3 z^{\prime} \sum_{l, l^{\prime}=1,2} \hat{\sigma}_{d_1 d_1}\left(z^{\prime}\right) \Delta\left(z-z^{\prime}\right) \hat{\sigma}_{d_{l^{\prime}} d_{l^{\prime}}}(z)+\delta_d\left(\hat{\sigma}_{d_1 d_1}+\hat{\sigma}_{d_2 d_2}\right)\right. \\
		\left.-g\left(\hat{\varepsilon}_1 \hat{\sigma}_{e_1 g_1}+\hat{\varepsilon}_2 \hat{\sigma}_{e_2 g_2}\right)-\Omega\left(\hat{\sigma}_{e_1 d_1}+\hat{\sigma}_{e_2 d_2}\right)+\text { c.c. }\right]
	\end{array}
	$$
	
	The Heisenberg-Langevin equations are given by: $\dot{\hat{\sigma}}_{\mu \nu}=-\frac{\imath}{\hbar}\left[\hat{\sigma}_{\mu \nu}, \hat{H}_{\text {int }}\right]$ where $\hat{\sigma}_{\mu \nu}=\frac{N}{L} \int d^3 z e^{-z_1^2 / w_a^2} \hat{\sigma}_{\mu \nu}(z)$ are the average density matrix elements. We now write all the Langevin equations derived above by denoting
	
	$\hat{\alpha}(z)=\int d^3 z^{\prime} e^{-\frac{z_1^2}{w_a^2}} \Delta_{l l^{\prime}}\left(z-z^{\prime}\right)\left(\hat{\sigma}_{d_1 d_1}+\hat{\sigma}_{d_2 d_2}\right)$. We also consider the ground state relaxation rate as $\gamma_c$ and the decay rate of excited Rydberg states as $\Gamma / 2$.
	
	\begin{gather*}
		\dot{\hat{\sigma}}_{g_1 e_1}=-i\left[\left(-\delta_1-i \frac{\Gamma}{2}\right) \hat{\sigma}_{g_1 e_1}+g \hat{\varepsilon}_1\left(\hat{\sigma}_{g_1 g_1}-\hat{\sigma}_{e_1 e_1}\right)+\Omega \hat{\sigma}_{g_1 d_1}\right]  \tag{2a}\\
		\hat{\sigma}_{g_2 e_2}=-i\left[\left(-\delta_2-i \frac{\Gamma}{2}\right) \hat{\sigma}_{g_2 e_2}+g \hat{\varepsilon}_2\left(\hat{\sigma}_{g_2 g_2}-\hat{\sigma}_{e_2 e_2}\right)+\Omega \hat{\sigma}_{g_2 d_2}\right]  \tag{2b}\\
		\hat{\sigma}_{e_1 d_1}=-i\left[\left(\hat{\alpha}+\delta_d+i \frac{\Gamma}{2}\right) \hat{\sigma}_{e_1 d_1}+g \hat{\varepsilon}_1^{\dagger} \hat{\sigma}_{g_1 d_1}-\Omega^{\dagger}\left(\hat{\sigma}_{e_1 e_1}-\hat{\sigma}_{d_1 d_1}\right)\right]  \tag{2c}\\
		\dot{\hat{\sigma}}_{e_2 d_2}=-i\left[\left(\hat{\alpha}+\delta_d+i \frac{\Gamma}{2}\right) \hat{\sigma}_{e_2 d_2}+g \hat{\varepsilon}_2^{\dagger} \hat{\sigma}_{g_2 d_2}-\Omega^{\dagger}\left(\hat{\sigma}_{e_2 e_2}-\hat{\sigma}_{d_2 d_2}\right)\right]  \tag{2d}\\
		\hat{\sigma}_{g_1 d_1}=-i\left[\left(-\delta_1+\hat{\alpha}+i \gamma_c\right) \hat{\sigma}_{g_1 d_1}+g \hat{\varepsilon}_1 \hat{\sigma}_{e_1 d_1}-\Omega^{\dagger} \hat{\sigma}_{g_1 e_1}+\delta_d \hat{\sigma}_{g_1 d_1}\right]  \tag{2e}\\
		\hat{\sigma}_{g_2 d_2}=-i\left[\left(-\delta_2+\hat{\alpha}+i \gamma_c\right) \hat{\sigma}_{g_2 d_2}+g \hat{\varepsilon}_2 \hat{\sigma}_{e_2 d_2}-\Omega^{\dagger} \hat{\sigma}_{g_2 e_2}+\delta_d \hat{\sigma}_{g_2 d_2}\right] \tag{2f}
	\end{gather*}
	
	Assuming weak probe fields $\left(\mathcal{E}_{1,2} \ll \Omega\right)$, equilibrium yields equally populated ground states ( $\hat{\sigma}_{g_1 g_1}=\hat{\sigma}_{g_2 g_2}=\frac{1}{2}$ ). Applying second-order perturbation, we expand the density matrix elements $\hat{\sigma}_{\mu \nu}$ for equations (2a), (2b), (2e), and (2f) with respect to Rabi frequency ratio $\epsilon=g \hat{\varepsilon}_{1,2} / \Omega$ and the adiabatic approximation ( $T_p^{-1} \ll \gamma_c$ ), under a strong control field $\left(|\Omega|^2 \gg \delta_{1,2}, \delta_d, \alpha\right)$ : time derivatives become negligible.
	
	\begin{gather*}
		\hat{\sigma}_{g l e_l}^{(2)}=\frac{-i}{\Omega^{+}}\left[\frac{\partial}{\partial t}-i\left(\delta_l-\delta_d-\alpha\right)+\gamma_c\right] \hat{\sigma}_{g l} d_l ; l=1,2  \tag{3a}\\
		\hat{\sigma}_{g l} \dot{d}_l^{(2)}=-\frac{g \hat{\varepsilon}_l}{2 \Omega}-\frac{g \hat{\varepsilon}_l}{2 \Omega}\left(\delta_l+\frac{i \Gamma}{2}\right) \frac{\left(\delta_l-\delta_d-\alpha\right)}{|\Omega|^2} ; l=1,2 \tag{3b}
	\end{gather*}
	
	The slowly varying propagation equation for weak quantum fields is \cite{fleischhauer2002}:
	
	\begin{equation*}
		\left(\frac{\partial}{\partial t}+c \frac{\partial}{\partial z}\right) \hat{\varepsilon}_l(z, t)=i g N\left(\frac{w}{w_a}\right)^2 \hat{\sigma}_{g l l_l} l=1,2 ; w=w_a w_f\left(w_a^2+w_f^2\right)^{-1 / 2} ; l=1,2 \tag{3c}
	\end{equation*}
	
	Substituting the value of $\hat{\sigma}_{g l e l}$ from Eq. (3a) and (3b) yield
	
	\begin{equation*}
		\left(\frac{\partial}{\partial t}+c \frac{\partial}{\partial z}\right) \hat{\varepsilon}_l(z, t)=-\frac{g^2 N}{2|\Omega|^2}\left(\frac{w}{w_a}\right)^2\left[\frac{\partial}{\partial t}-i\left(\delta_l-\delta_d-\alpha\right)+\gamma_c\right]\left[1+\left(\delta_l+\frac{i \Gamma}{2}\right) \frac{\left(\delta_l-\delta_d-\alpha\right)}{|\Omega|^2}\right] \hat{\varepsilon}_l ; l=1,2 \tag{4}
	\end{equation*}
	
	At resonance $\omega_{g l e_l}-\omega_p \approx 0, \delta_l=\omega_p-\omega_{g l e_l}-k_p v \mp \Delta \approx-k_p v \mp \Delta$. With $\Delta_d=\omega_d-\omega_{34}^0+\Delta^{\prime}$, $\delta_l-\delta_d=\left(k_d-k_p\right) v \pm\left(\Delta \mp \Delta_d\right)$. Since $\omega_p-\omega_d \ll \omega_{p, d},\left(k_d-k_p\right) v \approx 0$. The EIT $D-D$ phase shift conditions dictate $\delta \omega_{E I T}=\frac{|\Omega|^2}{\gamma_c \sqrt{k_0 L}}, \hat{\alpha}<\delta \omega_{E I T}, \alpha \ll|\Omega|^4 \Rightarrow \frac{\alpha}{|\Omega|^2} \approx 0, \gamma_c\left(k_p v+i \Gamma\right) \ll \Omega^2$, and $k_p v\left(\Delta+\Delta_d\right) \ll \Omega^2$.\cite{gasparoni2004} Rewriting uses $S_{1,2}=\frac{g^2 N}{2 \mathrm{c}|\Omega|^2}\left[1+\frac{\Delta\left(\Delta_d \pm \Delta\right)}{|\Omega|^2}\right], \kappa_{1,2}=\frac{g^2 N}{2 \mathrm{c}|\Omega|^2}\left[\gamma_c+\frac{\Gamma\left(\Delta_d \pm \Delta\right)}{|\Omega|^2}\right]$, and mixing angle $\tan ^2 \theta=\frac{g^2 N}{2|\Omega|^2}\left(w / w_a\right)^2$.
	
	\begin{equation*}
		\left[\frac{1}{c} \frac{\partial}{\partial t}+\frac{\partial}{\partial z}\right] \hat{\varepsilon}_l=-\tan ^2 \theta \frac{1}{c}\left[S_{1,2} \frac{\partial}{\partial t}+i\left(S_{1,2}\left(\Delta_d \pm \Delta\right)+\alpha\right)+\kappa_{1,2}\right] \hat{\varepsilon}_l ; l=1,2 \tag{5}
	\end{equation*}
	
	We define the dark state polariton (DSP) field, a quasiparticle representing the superposition of the atomic and cavity fields \cite{fleischhauer2002}, as $\stackrel{~}{\Psi}_l(z, t)=\sqrt{c / v_g} \hat{\varepsilon}_l(z, t)$, with $v_g=c \cos ^2 \theta$. For notational simplicity, we introduce $\zeta\left(\theta_l\right)=\left[\cos ^2 \theta_l+\sin ^2 \theta_l S_l\right]$, the effective group velocity $v_l=v_g / \zeta\left(\theta_l\right)$, the effective phase modulation coefficient $S_l^{\prime}=\frac{\sin ^2 \theta_0 s_l}{\zeta\left(\theta_l\right)}$, the effective linear absorption coefficient $\kappa_l^{\prime}=\frac{\sin ^2 \theta_l}{\zeta\left(\theta_1\right)}$, and the dipole-dipole (D-D) interaction-induced phase factor $\alpha^{\prime}=\frac{\sin ^2 \theta_\alpha \alpha}{\zeta\left(\theta_1\right)}$. By applying these definitions, Eq. (5) simplifies to
	
	\begin{equation*}
		\left[\frac{\partial}{\partial t}+v_l \frac{\partial}{\partial z}\right] \hat{\Psi}_l=-i\left[S_l^{\prime}\left(\Delta_d \pm \Delta\right)+\alpha^{\prime}\right] \hat{\Psi}_l-\kappa_l^{\prime} \hat{\Psi}_l ; l=1,2 \tag{6}
	\end{equation*}
	
	The solution to Eq. (6) is written as
	\begin{equation*}
		\hat{\Psi}_{l,out} = \hat{\Psi}_l(z \mp v_l t, 0) \exp(-\kappa_l' z) \exp(i\phi_{\pm}(z, t)z) , \phi_{1,2} = -\left( S_l'(\Delta_d \pm \Delta) + \frac{\sin^2 \theta_l \phi_{DD}}{\zeta(\theta_l)} \right) ; l=1,2 \tag{7}
	\end{equation*}
	
	Here, the nonlinear phase is given as $\phi_{DD} = \frac{1}{L} \int_0^\tau dt' \int_0^z dz' \Delta_{dd}(z - z') \left[\sin^2 \theta_l \hat{I}_l(z', t) + \sin^2 \theta_{l'} \hat{I}_{l'}(z', t)\right]$; $\hat{I}_l(z', t) = \hat{\Psi}_l^\dagger \hat{\Psi}_l = \hat{\sigma}_{d_l d_l}(z') = \hat{\sigma}_{d_l g}(z') \hat{\sigma}_{g d_l}(z')$ is the photon number operator of the weak probe. The integral $\iint d^2 z'' \Delta_{ij}(z - z'')$ can be evaluated as, where $w$ is the transverse Gaussian width of the pulse as
	
	\begin{equation*}
		\Delta_{dd}(z - z') = \iint d^2 z'' e^{-z_\perp^2 / w_a^2} \Delta_{ij}(z - z'') = \frac{2 C_{ij}}{\pi w^2 (\sqrt{2} w)^3} \left[2|\zeta| - \sqrt{\pi}(1 + 2\zeta^2)e^{\zeta^2} \text{erfc}(\zeta)\right] ; \zeta = (z - z') / \sqrt{2} w
	\end{equation*}
	
	The Eq. (7) is the central result of our work, which shows that the interacting oppositely circular-polarized DSPs acquire opposite linear phases $S_l'(\Delta \mp \Delta_d)$ owing to the Zeeman shift $\Delta_d$ and nonlinear phase shift $\frac{\sin^2 \theta_l \phi_{DD}}{\zeta(\theta_l)}$ owing to the D-D interaction. This feature allows us to generate a controlled phase difference between the left and right circular-polarized components of the DSPs, along with strong long-distance entanglement generation through the D-D interaction between both input probe pulses into the atomic medium.
	\subsection{The Routing Operator}
	\begin{figure}[htbp]
		\centering
		\includegraphics[width=0.9\linewidth]{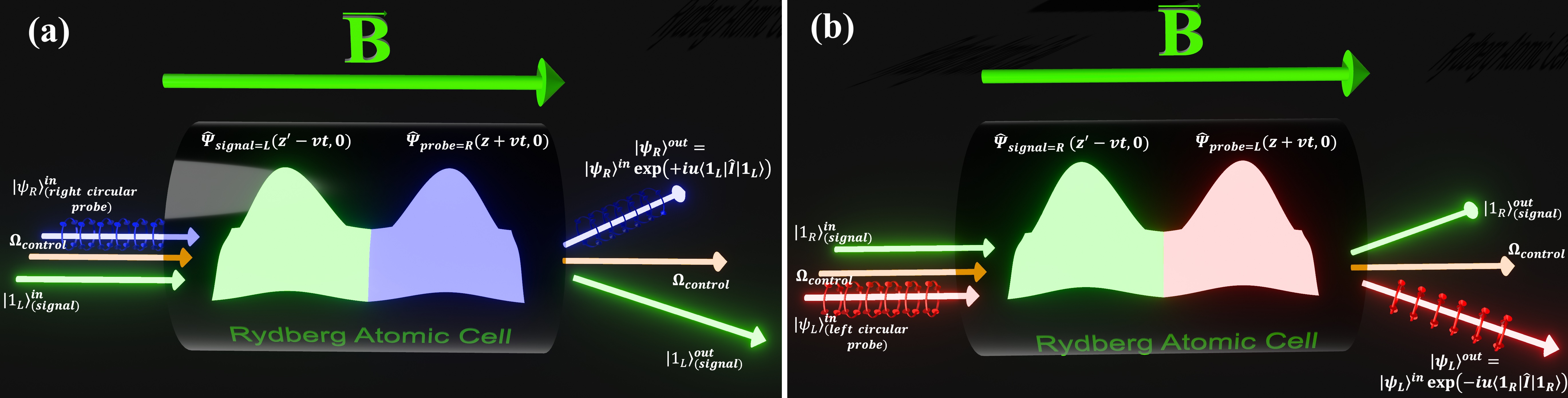}
		\caption{Routing operator's physical operation inside the Rydberg cell, showing interactions between counterpropagating weak ingle-photon "signal" and coherent "walker" probe pulses centred at $t=0$ around $z=0$ and $z=L$. (a) and (b) show left ( $\left|\psi_L\right\rangle$ ) and right $\left(\left|\psi_R\right\rangle\right)$ circularly polarized probes interacting with oppositely polarized signals, acquiring distinct linear (Zeeman shift) and nonlinear (D-D interaction) phases for routing.}
		\label{fig:fig2}
	\end{figure}
	To steer quantum walkers, a routing operator is defined to return oppositely polarized probes entangled with equal and opposite phases. We incorporate the effective linear phase into the polariton operator: $\hat{\Psi}_l(z \mp vt, 0) \to \hat{\Psi}_l(z \mp v_g t, 0) \exp(-\kappa_l' z) \exp(-iS_l'(\Delta_d \pm \Delta)z)$. Assuming equal mixing angles ($\theta_l \approx \theta$), EIT resonance $\frac{\Delta(\Delta \pm \Delta_d)}{|\Omega|^2} \simeq 0$, the phase factor simplifies to $\zeta(\theta_l) = \cos^2 \theta_l + \sin^2 \theta_l S_l' \simeq 1$.With the constants of motion $\hat{I}_j(t) = \hat{I}_j(0)$ and time integral $u = \sin^4 \theta \frac{1}{L} \int_0^T dt' \int_0^z dz' \Delta_{dd}(z - z')$, the polariton solution from Eq. (7) becomes:
	\begin{equation}
		\hat{\Psi}_l(z, t) = \hat{\Psi}_l(z, 0) \exp(-iu[\hat{I}_l(z, t) + \hat{I}_{l'}(z', t)]) , l=1,2. \tag{8}
	\end{equation}
	We consider two input pulses for the probe field operator (See Fig.~\ref{fig:fig2}): a coherent "walker" pulse $\hat{\Psi}_l(z, 0)|\psi_l\rangle = \hat{\Psi}_l(z \mp vt, 0) \exp(-iS_l'(\Delta_d \pm \Delta)z)|\psi_l\rangle = \varphi_l \exp(-iS_l'(\Delta_d \pm \Delta)z)|\psi_l\rangle = \Phi_l|\psi_l\rangle$ ,with eigenvalue, incorporating the linear phase shift $\Phi_l = \varphi_l \exp(-iS_l'(\Delta_d \pm \Delta)z)$ and a weak single-photon "signal" pulse state $|1_{l'}\rangle = \frac{1}{\sqrt{L}} \int dz f_{l'}(z) \hat{\Psi}_{l'}(z, 0)|0\rangle = \frac{1}{\sqrt{L}} \int dz f_{l'}(z) \hat{\Psi}_{l'}(z \mp vt, 0)\exp(-iS_l'(\Delta_d \pm \Delta)z)|0\rangle$, $F_{l'}(z) \equiv f_{l'}(z)\exp(-iS_l'(\Delta_d \pm \Delta)z)$ being the full spatial envelope, incorporating the linear phase satisfying the normalization $\frac{1}{L} \int dz |f_{l'}(z)|^2 = 1$. Operating on this input state $|\phi_l\rangle = |\psi_l\rangle \otimes |1_{l'}\rangle$ yields the following expectation value:
	\begin{equation}
		\langle 1_{l'}| \langle \psi_l| \hat{\Psi}_l(z, 0) |\psi_l\rangle |1_{l'}\rangle = \varphi_l(z \mp vt, 0) \exp(-iS_l'(\Delta_d \pm \Delta)z) \exp(-iu|f_{l'}(z')|^2) \tag{9}
	\end{equation}
	Neglecting the self-interaction term $u[\hat{I}_l(z, t)]$, which is negligible compared to the cross-interaction term, we consider pulses centred at $t = 0$ around $z = 0$ and $z = L$.
	Integrating via $vt' \to x$, $(x - z')/\sqrt{2}w \to \zeta'$, and $(L/\sqrt{2}w \to \infty)$ yields $\int_0^L dx \Delta_{dd}(x - z') = \sqrt{2}w \int_0^\infty d\zeta' \Delta_{dd}(\sqrt{2}w\zeta') = -\frac{C_{ij}}{\hbar w^2}$. Using normalization $\frac{1}{L} \int dz |f_{v'}(z)|^2 = 1$, the spatial integral evaluates to $\frac{1}{L} \int_0^T dt' \int_0^z dz' \Delta_{dd}((z + vt) - (z' - vt'))|f_k(z')|^2 = -i\frac{C_{tj}}{\hbar w^2 v}$.
	Approximating $S_l' \approx \sin^2 \theta$ (valid when $\frac{\Delta(\Delta \pm \Delta_d)}{|\Omega|^2} \simeq 0$), the expectation value becomes
	
	\begin{equation*}
		\langle 1_l| \langle \psi_l| \hat{\Psi}_l(L, L/v) |\psi_l \rangle |1_l \rangle = \varphi_l(0)\exp(\mp i\sin^2 \theta(\Delta \pm \Delta_d)L)\exp\left(-i\frac{C_{ij}\sin^4 \theta}{\hbar w^2 v}\right)
	\end{equation*}
	
	This contains the linear phase $(\Delta \pm \Delta_d)L$ and nonlinear phase $\frac{C_{ij} \sin^2 \theta}{\hbar w^2 v}$. For a cold atomic sample with $\pi$-polarized fields ($\omega_d \to \omega_{34}^0$), $\Delta_d = 0$ since $M_{F^v} = 0$. With $\Omega \gg g, \theta \to \frac{\pi}{2}$ yielding $\sin \theta \approx 1$. Thus, the routing operator $\hat{\Psi}\left(L, \frac{L}{v}\right)$ acts on probes $\varphi_L(0)$ and $\varphi_R(0)$ yields
	
	\begin{align}
		\langle \psi_L| \hat{\Psi}\left(L, \frac{L}{v}\right) |\psi_L \rangle &= \varphi_L(0) \exp \left( -i\left(\Delta L + \frac{C_{ij}}{\hbar w^2 v}\right) \right) \tag{11a} \\
		\langle \psi_R| \hat{\Psi}\left(L, \frac{L}{v}\right) |\psi_R \rangle &= \varphi_R(0) \exp \left( +i\left(\Delta L - \frac{C_{ij}}{\hbar w^2 v}\right) \right) \tag{11b}
	\end{align}
	
	Physically, the routing operator acting on the left and right circular probes supplies them with distinct phases, which can be used to allow them to route towards two distinct phase-selective waveguides. Finally, substituting the linear phase $\Delta L = \pm \frac{\pi}{2}$ and nonlinear phase $\frac{C_{ij}}{\hbar w^2 v} = \frac{\pi}{2}$ into eqs. (11a) and (11b) yields left and right circularly polarized output probes with equal and opposite resultant phases of $\mp 1$
	
	\begin{align}
		\langle \psi_L| \hat{\Psi}\left(L, \frac{L}{v}\right) |\psi_L \rangle &= \varphi_L(0) \exp \left( -i\left(\frac{\pi}{2} + \frac{\pi}{2}\right) \right) = -\varphi_L(0) \tag{12a} \\
		\langle \psi_R| \hat{\Psi}\left(L, \frac{L}{v}\right) |\psi_R \rangle &= \varphi_R(0) \exp \left( +i\left(\frac{\pi}{2} - \frac{\pi}{2}\right) \right) = +\varphi_R(0) \tag{12b}
	\end{align}
	
	The inverse operator $\hat{\Psi}^\dagger$ is defined to generate opposite phases as follows
	
	\begin{align}
		\langle \psi_L| \hat{\Psi}^\dagger\left(L, \frac{L}{v}\right) |\psi_L \rangle &= +\varphi_L(0) \tag{13a} \\
		\langle \psi_R| \hat{\Psi}^\dagger\left(L, \frac{L}{v}\right) |\psi_R \rangle &= -\varphi_R(0) \tag{13b}
	\end{align}
	\subsection{Implementation of the QRAM Architecture}
	\begin{figure}[htbp]
		\centering
		\includegraphics[width=0.9\linewidth]{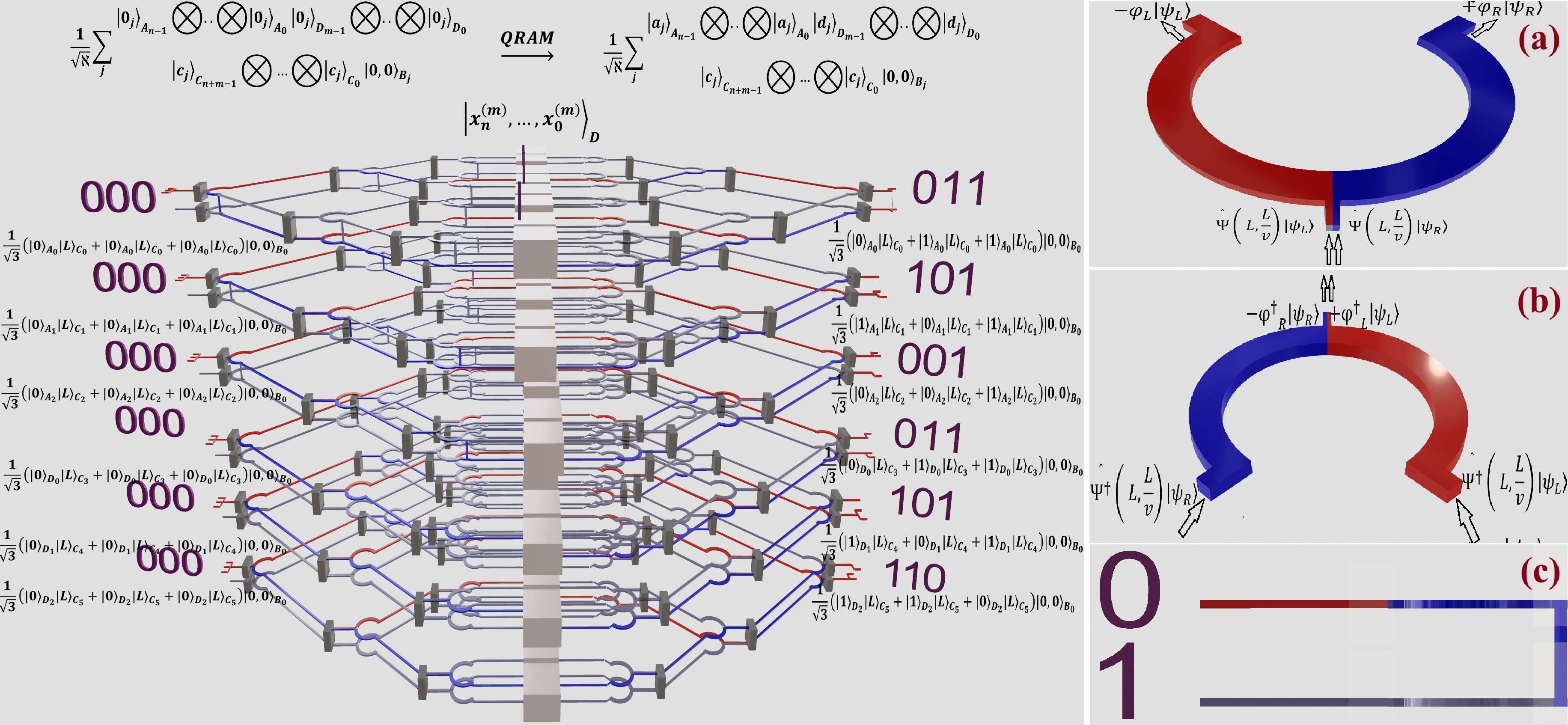}
		\caption{Depicting the Quantum Random Walk-based qRAM architecture, $n+m$ walkers traverse parallelized binary tree layers of interconnected phase-dependent waveguides, accessing memory addresses via double-rail encoding. (a) Routing operator $\underline{\Psi}$ exploits phase differences, steering left-polarized ( $|L\rangle$ ) and right-polarized ( $|R\rangle$ ) walkers into red and blue phase-selective waveguides. (b) Inverse operator $\Psi^{\dagger}$ enables deterministic retrieval. (c) The dual-rail signal mechanism.}
		\label{fig:fig3}
	\end{figure}
	Our architecture comprises a binary tree of interconnected phase-dependent cylindrical hollow-core photonic crystal waveguides that act as routing operators. Each node contains an ensemble of cold atoms excited to Rydberg states by strong control lasers in a longitudinal solenoidal magnetic field. Oppositely polarized probe pulses acquire opposite phases, acting as quantum walkers that navigate towards phase-selective waveguides (Fig.~\ref{fig:fig3}). The $n$-qubit walkers $\sum_j |0_j\rangle_{A_{n-1}} \otimes \cdots \otimes |0_j\rangle_{A_0}$ and $m$-qubit querying walkers $\sum_j |0_j\rangle_{D_{m-1}} \otimes \cdots \otimes |0_j\rangle_{D_0}$ retrieve superpositions $\frac{1}{\sqrt{N}}(|a_0\rangle_{A_{n-1}} \otimes \cdots \otimes |a_0\rangle_{A_0} + |a_1\rangle_{A_{n-1}} \otimes \cdots \otimes |a_1\rangle_{A_0} + \cdots + |a_j\rangle_{A_{n-1}} \otimes \cdots \otimes |a_j\rangle_{A_0})$ alongside corresponding data $\frac{1}{\sqrt{N}}(|d_0\rangle_{D_{m-1}} \otimes \cdots \otimes |d_0\rangle_{D_0} + |d_1\rangle_{D_{m-1}} \otimes \cdots \otimes |d_1\rangle_{D_0} + \cdots + |d_j\rangle_{D_{m-1}} \otimes \cdots \otimes |d_j\rangle_{D_0})$, where $a_i, d_i \in \{0,1\}\forall i$. Stored data is $|x_n^{(m)}, \dots, x_0^{(m)}\rangle_D$ . Simultaneously, $n+m$ walkers traverse $2(n+m)$ rails representing $|0\rangle/|1\rangle$, addressing $2^n$ memories. The walker's internal polarization state is $|c_j\rangle_{c_{n+m-1}} \otimes \cdots \otimes |c_j\rangle_{c_0}, c_j \in \{L, R\}$ denoting the left ($|L\rangle$) and right ($|R\rangle$) circular polarization. The collective position of all $n+m$ walkers routed among $j$ addresses is $|w, l\rangle_B = \sum_j |w, l\rangle_{B_j}$, requires $0 \leqslant l \leqslant 2^w - 1$ and $0 \leqslant w \leqslant n - 1$. The red and blue waveguides selectively transmit the $|L\rangle$ and $|R\rangle$ polarizations by virtue of their $\mp 1$ phases, respectively. This requires parametric pulse amplification via linear phase-matching \cite{zhao2022, shi2023} in integrated photonics such as AlGaAs or Si3N4 platforms for communication \cite{ledezma2022, ye2021, yan2022}. Ultimately, this qRAM coherently generates a unitary transformation that transports double-rail encoded information.
	
	\begin{align*}
		&\frac{1}{\sqrt{\aleph}} \sum_j |0_j\rangle_{A_{n-1}} \otimes .. \otimes |0_j\rangle_{A_0} |0_j\rangle_{D_{m-1}} \otimes .. \otimes |0_j\rangle_{D_0} |c_j\rangle_{C_{n+m-1}} \otimes \dots \otimes |c_j\rangle_{C_0} |0,0\rangle_{B_j} \\
		&\xrightarrow{QRAM} \frac{1}{\sqrt{\aleph}} \sum_j |a_j\rangle_{A_{n-1}} \otimes .. \otimes |a_j\rangle_{A_0} |d_j\rangle_{D_{m-1}} \otimes .. \otimes |d_j\rangle_{D_0} |c_j\rangle_{C_{n+m-1}} \otimes \dots \otimes |c_j\rangle_{C_0} |0,0\rangle_{B_j},
	\end{align*}
	
	Here, $\aleph$ is the normalisation constant.
	
	\subsubsection{Routing}
	\begin{figure}[htbp]
		\centering
		\includegraphics[width=\textwidth]{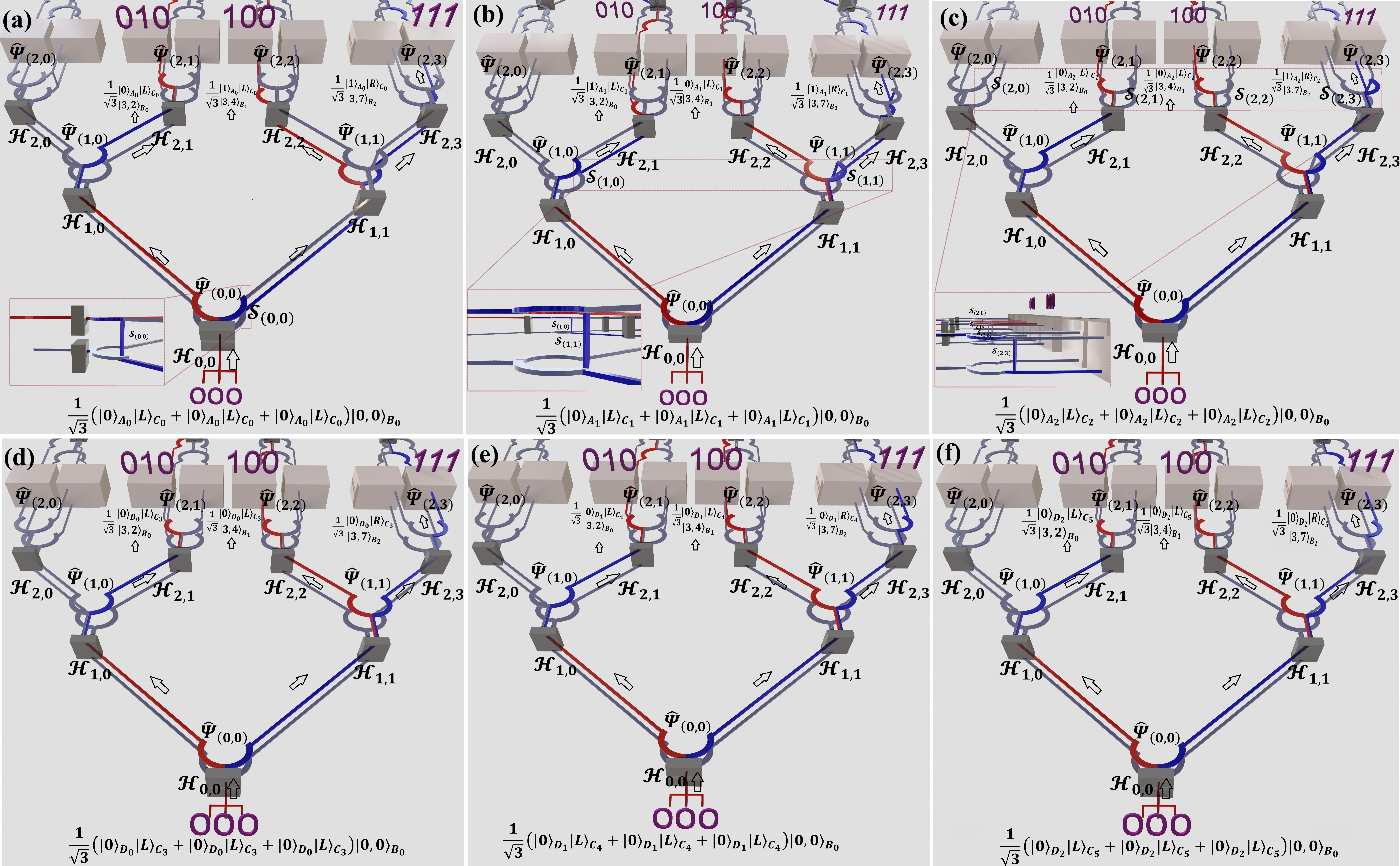}
		\caption{Maps transportation of the probe walkers navigating the binary tree from root $(0,0)$ to distinct memory locations $(010, 100, 111)$. Input in left circular polarization $|L\rangle$, walkers at node $(w,l)$ are manipulated by modified Hadamard $H(w,l)$ and routing $\Psi(w,l)$ operators to steer superpositions. Shift operator $S(w,l)$ concurrently encodes the address qubit. Panels (a)(c) and (d)-(f) show routing of address $|0\rangle_{A_2} \otimes |0\rangle_{A_1} \otimes |0\rangle_{A_0}$ and data $|0\rangle_{D_2} \otimes |0\rangle_{D_1} \otimes |0\rangle_{D_0}$ walkers, respectively. (Left (Right) circular polarization state probe $|L\rangle$ carrying waveguide color-coded as red (blue)).}
		\label{fig:fig4}
	\end{figure}
	Routing operators impart equal, opposite phases, directing walkers to phase-dependent waveguides. The $n$-level tree indexes nodes $( w, l )$, where level is $( 0 \leqslant w \leqslant n - 1 )$ and node is $l(0 \leqslant l \leqslant 2^w - 1)$. To propel the walkers towards the target addresses, requires a modified Hadamard operator. Before routing, all $n + m$ walkers are initialized at root (0,0) in left circular polarization $|L\rangle$ and dual-rail state $|0\rangle$:
	
	\begin{equation*}
		\frac{1}{\sqrt{\aleph}} \sum_j \left(|0_j\rangle_{A_{n-1}} \otimes .. \otimes |0_j\rangle_{A_0}\right) \left(|0_j\rangle_{D_{m-1}} \otimes .. \otimes |0_j\rangle_{D_0}\right) \left(|L_j\rangle_{c_{n+m-1}} \otimes ... \otimes |L_j\rangle_{c_0}\right) |0,0\rangle_{B_j}.
	\end{equation*}
	
	To simultaneously deliver walkers across $2(n + m)$ parallel binary tree sheets, Hadamard-like gates at each tree level route $l_{(w,l)}$ left and $r_{(w,l)}$ right circularly polarized walkers to child nodes $( w + 1, 2l )$ and $( w + 1, 2l + 1 )$, respectively. The gate \cite{asaka2023_2} on default left-circular walkers is defined as
	
	\begin{equation*}
		\mathcal{H}_{(w,l)} := \frac{1}{\sqrt{l_{(w,l)} + r_{(w,l)}}} \begin{pmatrix} \sqrt{l_{(w,l)}} & \sqrt{r_{(w,l)}} \\ \sqrt{r_{(w,l)}} & -\sqrt{l_{(w,l)}} \end{pmatrix}.
	\end{equation*}
	
	Equivalent to $e^{i\theta} R_y(\theta)R_z(\pi)$ with $\theta = 2\tan^{-1} \left(\sqrt{r_{(w,l)} / l_{(w,l)}}\right)$, it becomes the standard Hadamard when $l_{(w,l)} = r_{(w,l)} = 1$. It yields left and right walkers for the $i^{\text{th}}$ address/data walker:
	
	\begin{equation*}
		\sum_k |0_k\rangle_{A_i/D_i} \otimes |L_k\rangle_{C_i} \otimes |w, l\rangle_{B_k} \xrightarrow{\mathcal{H}_{(w,l)}} \left( \sum_k^{l_{(w,l)}} |0_k\rangle_{A_i/D_i} |L_k\rangle_{C_i} |w, l\rangle_{B_j} + \sum_k^{r_{(w,l)}} |0_k\rangle_{A_i/D_i} |R_k\rangle_{C_i} |w, l\rangle_{B_k} \right) ; \forall A_i/D_i
	\end{equation*}
	Finally, the routing operator $\Psi_{(w,l)}$ imparts opposite phases ($\mp 1$), predictably steering $|L\rangle$ (red) and $|R\rangle$ (blue) walkers to the left and right daughter nodes. For the $i^{\text{th}}$ address/data walker, the operation is given as
	
	\begin{align*}
		&\left( \sum_k^{l_{(w,\ell)}} |0_k\rangle_{A_i/D_i} |L_k\rangle_{C_i} + \sum_k^{r_{(w,\ell)}} |0_k\rangle_{A_i/D_i} |R_k\rangle_{C_i} \right) |w, \ell\rangle_{B_k} \\
		&\xrightarrow{\Psi_{(w,\ell)}} \left( \sum_k^{l_{(w,\ell)}} (-1)^{l_{(w,\ell)}} \times |0_k\rangle_{A_i/D_i} |L_k\rangle_{C_i} |w + 1, 2\ell\rangle_{B_k} + \sum_k^{r_{(w,\ell)}} |0_k\rangle_{A_i/D_i} |R_k\rangle_{C_i} |w + 1, 2\ell + 1\rangle_{B_k} \right) ; \forall A_i, D_i, C_i
	\end{align*}
	
	Following routing, the address qubits are double-rail encoded onto each walker. The shift operator $S_{(w,l)}$ shifts the right-circularly polarized (blue) walker into the dual-rail state $|1\rangle$, preserving the left circularly polarized walker(red) in state $|0\rangle$. Consequently, for all $j$ walkers, the internal polarization $|L\rangle$ ($|R\rangle$) transitions the most significant qubit to $|0\rangle (|1\rangle)$. This utilizes polarization-sensitive waveguides \cite{xie2020, zheng2023}. To copy the address information, the $i^{\text{th}}$ shift operator $S_{(i,l)}$ acts on the $i^{\text{th}}$ walker at the $i^{\text{th}}$ level. Based on polarization $|L(R)\rangle_{c_c} |i, l\rangle_B$, its path shifts to the dual-rail state $|0(1)\rangle_{A_i}^{(i,l)}$. The shifting operation on the $i^{\text{th}}$ address state walker is summarized as:
	
	\begin{align*}
		&\sum_k^{l_{(w,\ell)}} (-1)^{l_{(w,\ell)}} \times |0_k\rangle_{A_i} |L_k\rangle_{C_i} |i + 1, 2\ell\rangle_{B_j} + \sum_k^{r_{(w,\ell)}} |0_k\rangle_{A_i} |R_k\rangle_{C_i} |i + 1, 2\ell + 1\rangle_{B_k} \\
		&\xrightarrow{S_{(i,\ell)}} \sum_k^{l_{(w,\ell)}} (-1)^{l_{(w,\ell)}} \times |0_k\rangle_{A_i} |L_k\rangle_{C_i} |i + 1, 2\ell\rangle_{B_j} + \sum_k^{r_{(w,\ell)}} |1_k\rangle_{A_i} |R_k\rangle_{C_i} |i + 1, 2\ell + 1\rangle_{B_k} ; \forall A_i, D_i, C_i
	\end{align*}
	
	The entire routing operation of $\mathcal{R}_{(w+1|w)}^i$ from node $(w, l)$ to daughter nodes $(w + 1, 2l)$ and $(w + 1, 2l + 1)$ for the $i^{\text{th}}$ walker can be summarized as
	\begin{align*}
		\mathcal{R}_{(w+1|w)}^i : \sum_k |0_k\rangle_{A_i} |0_k\rangle_{D_i} |L_k\rangle_{C_i} |w, \ell\rangle_{B_k} &\to \sum_k^{l_{(w,\ell)}} (-1)^{l_{(w,\ell)}} \times |0_k\rangle_{A_i} |0_k\rangle_{D_i} |L_k\rangle_{c_i} |i+1, 2\ell\rangle_{B_j} \\
		&\quad + \sum_k^{r_{(w,\ell)}} |1_k\rangle_{A_i} |0_k\rangle_{D_i} |R_k\rangle_{c_i} |i+1, 2\ell+1\rangle_{B_k} ; \forall A_i, D_i, C_i; \\
		\mathcal{R}_{(w+1|w)}^i &= \sum_{\ell=0}^{2^w-1} \mathcal{S}_{(i,\ell)} \hat{\Psi}_{(w,\ell)} \mathcal{H}_{(w,\ell)}
	\end{align*}
	
	Therefore, the entire routing operation $\mathcal{R}$ of the $n + m$ walkers from node $(0,0)$ to the $n$-qubit address state will be given by applying the tensor product of $\mathcal{R}_{(w+1|w)}^i$ recursively for each level of the binary tree to all $n + m$ walkers
	
	\begin{align*}
		\mathcal{R} : \frac{1}{\sqrt{\aleph}} &\left( \sum_k |0_k\rangle_{A_{n-1}} \otimes .. \otimes |0_k\rangle_{A_0} |0_k\rangle_{D_{m-1}} \otimes .. \otimes |0_k\rangle_{D_0} |L_k\rangle_{c_{n+m-1}} \otimes \dots \otimes |L_k\rangle_{c_0} |0,0\rangle_{B_j} \right) \\
		&\to \frac{1}{\sqrt{\aleph}} \left( \sum_k^j |a_k\rangle_{A_{n-1}} \otimes .. \otimes |a_k\rangle_{A_0} |0_k\rangle_{D_{m-1}} \otimes .. \otimes |0_k\rangle_{D_0} |c_k\rangle_{c_{n+m-1}} \otimes \dots \otimes |c_k\rangle_{c_0} |n,k\rangle_{B_k} \right) ; \\
		&\mathcal{R} = \bigotimes_{i=0}^{n+m-1} \mathcal{R}_{(n|n-1)}^i \mathcal{R}_{(n-1|n-2)}^i \dots \mathcal{R}_{(2|1)}^i \mathcal{R}_{(1|0)}^i
	\end{align*}
	
	To illustrate, consider a qRAM operation retrieving a superposition of three-qubit address state walkers, $\frac{1}{\sqrt{3}}(|0\rangle_{A_0} \otimes |1\rangle_{A_1} \otimes |0\rangle_{A_2} + |1\rangle_{A_0} \otimes |0\rangle_{A_1} \otimes |0\rangle_{A_2} + |1\rangle_{A_0} \otimes |1\rangle_{A_1} \otimes |1\rangle_{A_2})$, and data walkers, $\frac{1}{\sqrt{3}}(|0\rangle_{D_0} \otimes |1\rangle_{D_1} \otimes |1\rangle_{D_2} + |1\rangle_{D_0} \otimes |0\rangle_{D_1} \otimes |1\rangle_{D_2} + |1\rangle_{D_0} \otimes |1\rangle_{D_1} \otimes |0\rangle_{D_2})$ (See Fig.~\ref{fig:fig4}). All six input walkers initialize as $\sum_{k=0}^2 \frac{1}{\sqrt{3}} (|0\rangle_{A_0} \otimes |0\rangle_{A_1} \otimes |0\rangle_{A_2})(|0\rangle_{D_0} \otimes |0\rangle_{D_1} \otimes |0\rangle_{D_2})(|L\rangle_{c_0} \otimes .. \otimes |L\rangle_{c_5})|0,0\rangle_{B_k}$. Routing the address walkers at root node $(0,0)$ uses Hadamard $\mathcal{H}_{(0,0)}$ and routing $\hat{\Psi}_{(0,0)}$ operators. This separates components: one $|\psi_L\rangle$ steers to node $(1,0)$ via the red waveguide (+1 phase), while two $|\psi_R\rangle$ direct to node $(1,1)$ via the blue waveguide (-1 phase).
	
	\begin{align*}
		&\frac{1}{\sqrt{3}} \Big( (|0\rangle_{A_0} \otimes |0\rangle_{A_1} \otimes |0\rangle_{A_2})(|L\rangle_{C_0} \otimes .. \otimes |L\rangle_{C_5}) + (|0\rangle_{A_0} \otimes |0\rangle_{A_1} \otimes |0\rangle_{A_2})(|L\rangle_{C_0} \otimes .. \otimes |L\rangle_{C_5}) \\
		&\qquad + (|0\rangle_{A_0} \otimes |0\rangle_{A_1} \otimes |0\rangle_{A_2})(|L\rangle_{C_0} \otimes .. \otimes |L\rangle_{C_5}) \Big) (|0\rangle_{D_0} \otimes |0\rangle_{D_1} \otimes |0\rangle_{D_2}) |0,0\rangle_{B_0} \xrightarrow{\mathcal{R}_{(1|0)}^2 = \mathcal{H}_{(0,0)} \Psi_{(0,0)} \mathcal{S}_{(0,0)}} \\
		&\left( \frac{1}{\sqrt{3}} (-1) (|0\rangle_{A_0} \otimes |0\rangle_{A_1} \otimes |0\rangle_{A_2}) (|L\rangle_{C_0} \otimes .. \otimes |L\rangle_{C_5}) |1,0\rangle_{B_0} \right. \\
		&\quad \left. + \frac{\sqrt{2}}{\sqrt{3}} \Big( |1\rangle_{A_0} \otimes |0\rangle_{A_1} \otimes |0\rangle_{A_2} + |1\rangle_{A_0} \otimes |0\rangle_{A_1} \otimes |0\rangle_{A_2} \Big) \right) (|R\rangle_{C_0} \otimes .. \otimes |R\rangle_{C_5}) \otimes (|0\rangle_{D_0} \otimes |0\rangle_{D_1} \otimes |0\rangle_{D_2}) |1,1\rangle_{B_1} ; \\
		&\mathcal{H}_{(0,0)} = \frac{1}{\sqrt{3}} \begin{bmatrix} 1 & \sqrt{2} \\ \sqrt{2} & -1 \end{bmatrix}
	\end{align*}
	
	Next, we apply the operators $\mathcal{H}_{(1,0)} \hat{\Psi}_{(1,0)}$ and $\mathcal{H}_{(1,1)} \hat{\Psi}_{(1,1)}$ on the respective daughter states to obtain
	
	\begin{align*}
		&\left( \frac{1}{\sqrt{3}} (-1) (|0\rangle_{A_0} \otimes |0\rangle_{A_1} \otimes |0\rangle_{A_2}) (|L\rangle_{C_0} \otimes .. \otimes |L\rangle_{C_5}) |1,0\rangle_{B_0} \right. \\
		&\quad \left. + \frac{\sqrt{2}}{\sqrt{3}} \Big( |1\rangle_{A_0} \otimes |0\rangle_{A_1} \otimes |0\rangle_{A_2} + |1\rangle_{A_0} \otimes |0\rangle_{A_1} \otimes |0\rangle_{A_2} \Big) \right) (|R\rangle_{C_0} \otimes .. \otimes |R\rangle_{C_5}) \otimes (|0\rangle_{D_0} \otimes |0\rangle_{D_1} \otimes |0\rangle_{D_2}) |1,1\rangle_{B_1} \\
		&\xrightarrow{\mathcal{R}_{(2|1)}^2 = \mathcal{H}_{(1,0)} \Psi_{(1,0)} \mathcal{S}_{(1,0)} + \mathcal{H}_{(1,1)} \Psi_{(1,1)} \mathcal{S}_{(1,1)}} \\
		&\frac{1}{\sqrt{3}} \Big( (-1)(|0\rangle_{A_0} \otimes |1\rangle_{A_1} \otimes |0\rangle_{A_2}) (|R\rangle_{C_0} \otimes .. \otimes |R\rangle_{C_5}) |2,1\rangle_{B_0} + (-1)(|1\rangle_{A_0} \otimes |0\rangle_{A_1} \otimes |0\rangle_{A_2}) (|L\rangle_{C_0} \otimes .. \otimes |L\rangle_{C_5}) |2,2\rangle_{B_1} \\
		&\qquad + (|1\rangle_{A_0} \otimes |1\rangle_{A_1} \otimes |0\rangle_{A_2}) (|R\rangle_{C_0} \otimes .. \otimes |R\rangle_{C_5}) |2,3\rangle_{B_1} \Big) \otimes (|0\rangle_{D_0} \otimes |0\rangle_{D_1} \otimes |0\rangle_{D_2}) ; \\
		&\mathcal{H}_{(1,0)} = \begin{bmatrix} 0 & 1 \\ 1 & 0 \end{bmatrix}, \mathcal{H}_{(1,1)} = \frac{1}{\sqrt{2}} \begin{bmatrix} 1 & 1 \\ 1 & -1 \end{bmatrix}
	\end{align*}
	
	Applying the operators $\mathcal{H}_{(2,1)} \hat{\Psi}_{(2,1)}, \mathcal{H}_{(2,2)} \hat{\Psi}_{(2,2)}$, and $\mathcal{H}_{(2,3)} \hat{\Psi}_{(2,3)}$ on the respective daughter states and routes them to memory addresses (101), (100), and (111). Additionally, the $i^{\text{th}}$ level shift operator $S_{(i,l)}$ acts on the $i\text{th}$ state, modifying the address walker's dual-rail state $|0\rangle_{A_i} \to |1\rangle_{A_i}$ for $|L\rangle_{c_i}(|R\rangle_{c_i})$.
	
	\begin{align*}
		&\frac{1}{\sqrt{3}} \Big( (-1)(|0\rangle_{A_0} \otimes |1\rangle_{A_1} \otimes |0\rangle_{A_2})(|R\rangle_{C_0} \otimes .. \otimes |R\rangle_{C_5})|2,1\rangle_{B_0} + (-1)(|1\rangle_{A_0} \otimes |0\rangle_{A_1} \otimes |0\rangle_{A_2})(|L\rangle_{c_0} \otimes .. \otimes |L\rangle_{c_5})|2,2\rangle_{B_1} \\
		&\qquad + (|1\rangle_{A_0} \otimes |1\rangle_{A_1} \otimes |0\rangle_{A_2})(|R\rangle_{C_0} \otimes .. \otimes |R\rangle_{C_5})|2,3\rangle_{B_1} \Big) \otimes (|0\rangle_{D_0} \otimes |0\rangle_{D_1} \otimes |0\rangle_{D_2}) \\
		&\xrightarrow{\mathcal{R}_{(3|2)}^2 = \mathcal{H}_{(2,1)}\Psi_{(2,1)}\mathcal{S}_{(2,1)} + \mathcal{H}_{(2,2)}\Psi_{(2,2)}\mathcal{S}_{(2,2)} + \mathcal{H}_{(2,3)}\Psi_{(2,3)}\mathcal{S}_{(2,3)}} \\
		&\frac{1}{\sqrt{3}} \Big( (-1)^2(|0\rangle_{A_0} \otimes |1\rangle_{A_1} \otimes |0\rangle_{A_2})(|L\rangle_{C_0} \otimes .. \otimes |L\rangle_{C_5})|3,2\rangle_{B_0} + (-1)^2(|1\rangle_{A_0} \otimes |0\rangle_{A_1} \otimes |0\rangle_{A_2})(|L\rangle_{c_0} \otimes .. \otimes |L\rangle_{C_5})|3,4\rangle_{B_1} \\
		&\qquad + (|1\rangle_{A_0} \otimes |1\rangle_{A_1} \otimes |1\rangle_{A_2})(|R\rangle_{c_0} \otimes .. \otimes |R\rangle_{C_5})|3,7\rangle_{B_1} \Big) \otimes (|0\rangle_{D_0} \otimes |0\rangle_{D_1} \otimes |0\rangle_{D_2}) ; \\
		&\mathcal{H}_{(2,1)} = \begin{bmatrix} 0 & 1 \\ 1 & 0 \end{bmatrix}, \mathcal{H}_{(2,2)} = \begin{bmatrix} 1 & 0 \\ 0 & -1 \end{bmatrix}, \mathcal{H}_{(2,3)} = \begin{bmatrix} 1 & 0 \\ 0 & -1 \end{bmatrix}
	\end{align*}
	
	As required, we obtain the address state walkers encoded with the desired address state information of the memory addresses $|010\rangle$, $|100\rangle$, \textit{and} $|111\rangle$ after the routing.

	\subsubsection{Querying}
	
	\begin{figure}[htbp]
		\centering
		\includegraphics[width=\textwidth]{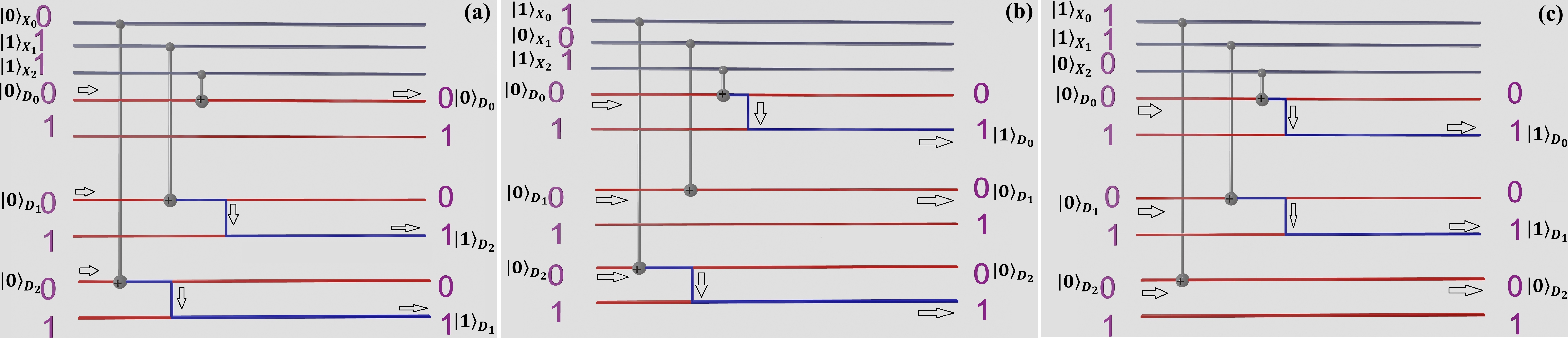}
		\caption{Details the qRAM querying phase using dual-rail CNOT operations. Target memory information acts as control bits $x_0, x_1$, and $x_2$, modifying data walkers $D_0, D_1$, and $D_2$ stored in waveguides ($|L\rangle$). Successive CNOT gates copy target data superpositions. Panels (a)-(c) show quantum querying from locations (010), (100), (111), respectively.}
		\label{fig:fig5}
	\end{figure}
	After bringing the $n + m$ probe walkers $\frac{1}{\sqrt{\aleph}} \sum_j \left(|0_j\rangle_{A_{m-1}} \otimes .. \otimes |0_j\rangle_{A_0}\right) \left(|0_j\rangle_{D_{m-1}} \otimes .. \otimes |0_j\rangle_{D_0}\right)$ with internal polarization states $|c_j\rangle_{c_{n+m-1}} \otimes \dots \otimes |c_j\rangle_{c_0}$ to memory addresses, we copy the $m$-qubit information $x_j^{(m)}$ into $m$ data walkers via dual-rail encoding using a CNOT operation. Copying the $m$-qubit data $|d_i\rangle_{x_{m-1}} \otimes \dots \otimes |d_i\rangle_{x_0}$ of the $i^{\text{th}}$ cell onto the $j^{\text{th}}$ component $|0_j\rangle_{D_{m-1}} \otimes .. \otimes |0_j\rangle_{D_0}$ is given by:
	\begin{equation*}
		\frac{1}{\sqrt{\aleph}} \sum_j \left( |d_i\rangle_{x_{m-1}} |0_j\rangle_{D_{m-1}} \otimes \dots \otimes |d_i\rangle_{x_0} |0_j\rangle_{D_0} \right) \xrightarrow{\text{CNOT}} \frac{1}{\sqrt{\aleph}} \sum_j |d_i\rangle_{D_{m-1}} \otimes .. \otimes |d_i\rangle_{D_0}
	\end{equation*}
	Thus, after the querying operation, we have
	\begin{align*}
		&\frac{1}{\sqrt{\aleph}} \sum_k \left(|a_k\rangle_{A_{n-1}} \otimes .. \otimes |a_k\rangle_{A_0}\right) \left(|0_k\rangle_{D_{m-1}} \otimes .. \otimes |0_k\rangle_{D_0}\right) \left(|c_k\rangle_{c_{n+m-1}} \otimes \dots \otimes |c_k\rangle_{c_0}\right) |0,0\rangle_{B_j} \\
		&\to \frac{1}{\sqrt{\aleph}} \sum_k^j \left(|a_k\rangle_{A_{n-1}} \otimes .. \otimes |a_k\rangle_{A_0}\right) \left(|d_k\rangle_{D_{m-1}} \otimes .. \otimes |d_k\rangle_{D_0}\right) \left(|c_k\rangle_{c_{n+m-1}} \otimes \dots \otimes |c_k\rangle_{c_0}\right) |n,k\rangle_{B_k}
	\end{align*}
	
	Continuing our qRAM example across six layers of three-level binary trees, we aim to retrieve a superposition of three-qubit address state walkers, $s_0|0\rangle_{A_0} \otimes |1\rangle_{A_1} \otimes |0\rangle_{A_2} + s_1|1\rangle_{A_0} \otimes |0\rangle_{A_1} \otimes |0\rangle_{A_2} + s_2|1\rangle_{A_0} \otimes |1\rangle_{A_1} \otimes |1\rangle_{A_2}$, and three-qubit data state walkers, $s_0|0\rangle_{D_0} \otimes |1\rangle_{D_1} \otimes |1\rangle_{D_2} + s_1|1\rangle_{D_0} \otimes |0\rangle_{D_1} \otimes |1\rangle_{D_2} + s_2|1\rangle_{D_0} \otimes |1\rangle_{D_1} \otimes |0\rangle_{D_2}$. At the target memory locations, the combined state of the six address and data walkers is
	\begin{align*}
		&\frac{1}{\sqrt{3}} \left((|0\rangle_{A_0} \otimes |1\rangle_{A_1} \otimes |0\rangle_{A_2})(|0\rangle_{D_0} \otimes |0\rangle_{D_1} \otimes |0\rangle_{D_2})(|L\rangle_{c_0} \otimes \dots \otimes |L\rangle_{c_5})\right) + \\
		&\frac{1}{\sqrt{3}} \left((|1\rangle_{A_0} \otimes |0\rangle_{A_1} \otimes |0\rangle_{A_2})(|0\rangle_{D_0} \otimes |0\rangle_{D_1} \otimes |0\rangle_{D_2})(|L\rangle_{c_0} \otimes \dots \otimes |L\rangle_{c_5})\right) + \\
		&\frac{1}{\sqrt{3}} \left((|1\rangle_{A_0} \otimes |1\rangle_{A_1} \otimes |1\rangle_{A_2})(|0\rangle_{D_0} \otimes |0\rangle_{D_1} \otimes |0\rangle_{D_2})(|R\rangle_{c_0} \otimes \dots \otimes |R\rangle_{c_5})\right)
	\end{align*}
	
	To copy the superposition of the data qubit, $\frac{1}{\sqrt{3}} \left(|0\rangle_{D_0} \otimes |0\rangle_{D_1} \otimes |0\rangle_{D_2} + |0\rangle_{D_0} \otimes |0\rangle_{D_1} \otimes |0\rangle_{D_2} + |0\rangle_{D_0} \otimes |0\rangle_{D_1} \otimes |0\rangle_{D_2}\right)$, from addresses $(010)$, $(100)$, and $(111)$, we apply a CNOT operation. (See Fig.~\ref{fig:fig5}) We utilize the memory information $|x^{(3)}\rangle_{(010)} = |0\rangle_{x_0} \otimes |1\rangle_{x_1} \otimes |1\rangle_{x_2}$, $|x^{(3)}\rangle_{(100)} = |1\rangle_{x_0} \otimes |0\rangle_{x_1} \otimes |1\rangle_{x_2}$, and $|x^{(3)}\rangle_{(111)} = |1\rangle_{x_0} \otimes |1\rangle_{x_1} \otimes |0\rangle_{x_2}$ as controls acting on the data state walker. This successfully copies the qubit onto the data walker states as follows:
	
	\begin{align*}
		&\frac{1}{\sqrt{3}} \Big( |0\rangle_{X_0} \otimes |1\rangle_{X_1} \otimes |1\rangle_{X_2}(|0\rangle_{D_0} \otimes |0\rangle_{D_1} \otimes |0\rangle_{D_2}) + |1\rangle_{X_0} \otimes |0\rangle_{X_1} \otimes |1\rangle_{X_2}(|0\rangle_{D_0} \otimes |0\rangle_{D_1} \otimes |0\rangle_{D_2}) + \\
		&|1\rangle_{X_0} \otimes |1\rangle_{X_1} \otimes |0\rangle_{X_2}(|0\rangle_{D_0} \otimes |0\rangle_{D_1} \otimes |0\rangle_{D_2}) \Big) \xrightarrow{\text{CNOT}} \frac{1}{\sqrt{3}} \Big( |0\rangle_{D_0} \otimes |1\rangle_{D_1} \otimes |1\rangle_{D_2} + |1\rangle_{D_0} \otimes |0\rangle_{D_1} \otimes |1\rangle_{D_2} + |1\rangle_{D_0} \otimes |1\rangle_{D_1} \otimes |0\rangle_{D_2} \Big)
	\end{align*}
	
	Therefore, we successfully obtained the desired data state information from the given memory addresses $|010\rangle$, $|100\rangle$, \textit{and} $|111\rangle$ after the query.
	
	\subsubsection{Retrieval}
	\begin{figure}[htbp]
		\centering
		\includegraphics[width=\textwidth]{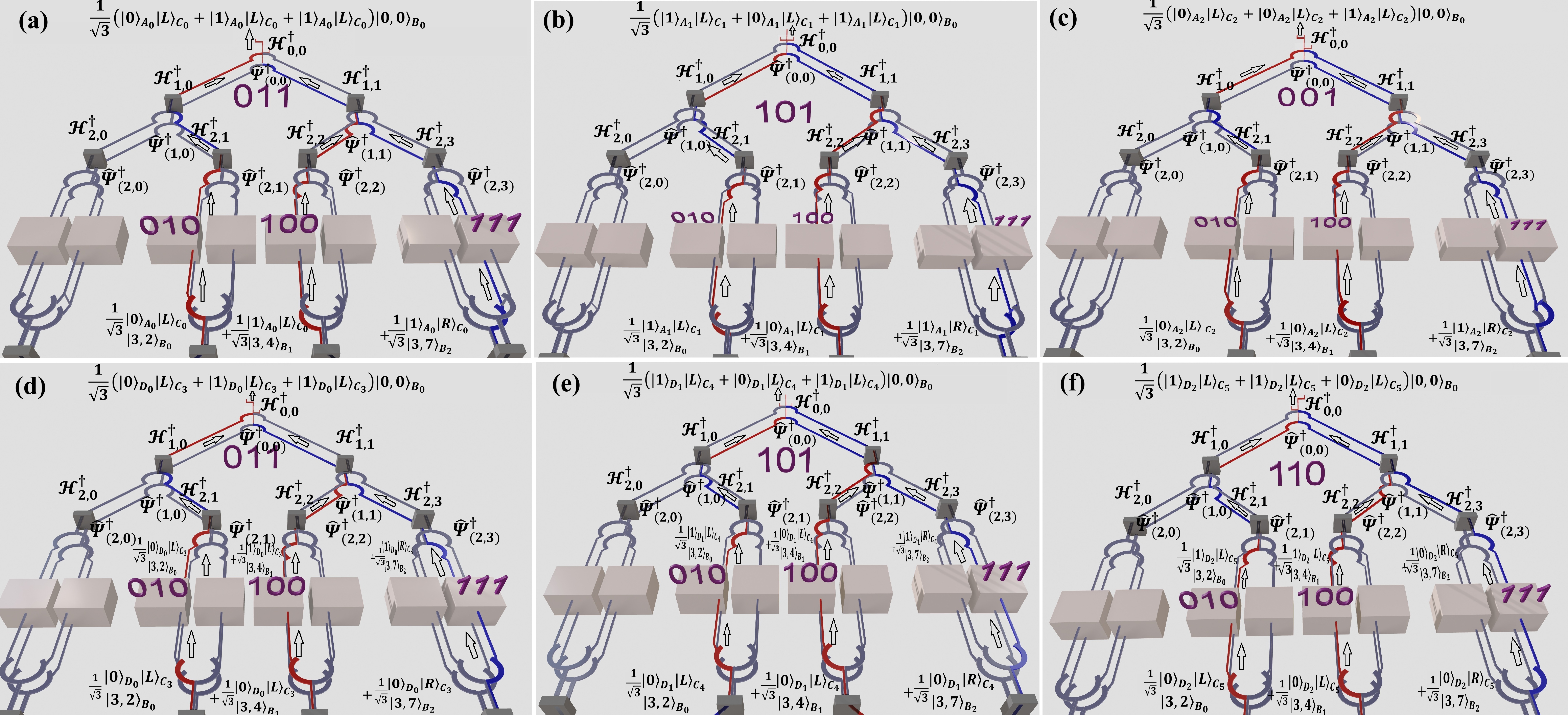}
		\caption{Tracing deterministic retrieval, returning quantum walkers move from queried memories in reverse through binary tree layers back to root node. Walkers interact with conjugate routing $\Psi(w,l)^\dagger$ and modified Hadamard $H(w,l)^\dagger$ operators, neutralizing phase differences and resetting polarizations to default state $|L\rangle$. This ensures successful coalescence of address and data walkers at root $(0,0)$. Panels (a)-(c) and (d)-(f) display this complete retrieval cycle visually. (Left (Right) circular polarization state probe $|L\rangle$ carrying waveguide color-coded as red (blue)).}
		\label{fig:fig6}
	\end{figure}
	Retrieval returns $n + m$ walkers with encoded information $\frac{1}{\sqrt{N}} \sum_k^j (|a_k\rangle_{A_{n-1}} \otimes \cdots \otimes |a_k\rangle_{A_0})(|d_k\rangle_{D_{m-1}} \otimes \cdots \otimes |d_k\rangle_{D_0})(|c_k\rangle_{c_{n+m-1}} \otimes \cdots \otimes |c_k\rangle_{c_0})|n, k\rangle_{B_k}$ through inverse tree layers via operators $\hat{\Psi}_{(w,\ell)}^\dagger$ and $\mathcal{H}_{(w,\ell)}^\dagger$. At node $(w, \ell)$, $\hat{\Psi}_{(w,\ell)}^\dagger$ imparts opposite $(\pm 1)$ phases to $|L\rangle$ and $|R\rangle$ walkers returning from daughter nodes $(w + 1, \ell \bmod 2)$ and $(w + 1, \ell \bmod 2 + 1)$, neutralizing their relative phase difference. For the $i^{\text{th}}$ address/data walker $A_i/D_i$ carrying the $a_k/d_k$ qubit, this operation is expressed as:
	
	\begin{align*}
		&\frac{1}{\sqrt{\aleph}} \left( \sum_k^{r_{(w+1,\ell)}} (-1)^{l_{(w,\ell)}} \times |a_k/d_k\rangle_{A_i/D_i} |R_k\rangle_{c_i} |w + 1, \ell \bmod 2\rangle_{B_k} + \sum_k^{l_{(w+1,\ell)}} |a_k/d_k\rangle_{A_i/D_i} |L_k\rangle_{C_i} |w + 1, \ell \bmod 2 + 1\rangle_{B_k} \right) \\
		&\xrightarrow{\Psi_{(w,\ell)}^\dagger} \frac{1}{\sqrt{N}} \left( \sum_k^{l_{(w+1,\ell)}} |a_k/d_k\rangle_{A_i/D_i} |R_k\rangle_{c_i} + \sum_k^{r_{(w+1,\ell)}} |a_k/d_k\rangle_{A_i/D_i} |L_k\rangle_{c_i} \right) |w, \ell\rangle_{B_k} ; \forall \frac{A_i}{D_i}
	\end{align*}
	
	Subsequently, the unitary modified Hadamard operator $\mathcal{H}_{(w, \ell \bmod 2)}^\dagger (\mathcal{H}_{(w, \ell \bmod 2 + 1)}^\dagger)$ at even(odd) node of level $w$ resets the polarization of the left and right circularly polarized walkers to left(right) circular polarization. This restores the coherent collection of walkers propagating towards the next parent node $(w - 1, \ell)$ via red and blue phase-sensitive waveguides. For the $i^{\text{th}}$ address/data walker $A_i/D_i$ carrying the $a_k$ state qubit, the operation is given as
	\begin{align*}
		&\frac{1}{\sqrt{\aleph}} \left( \sum_k^{r_{(w+1,\ell)}} |a_k/d_k\rangle_{A_i/D_i} |R_k\rangle_{c_i} + \sum_k^{l_{(w+1,\ell)}} |a_k/d_k\rangle_{A_i/D_i} |L_k\rangle_{c_i} \right) |w, \ell \bmod 2(\ell \bmod 2 + 1)\rangle_{B_k} \\
		&\xrightarrow{\mathcal{H}_{(w,\ell)}^\dagger} \frac{1}{\sqrt{\aleph}} \left( \sum_k^{l_{(w,\ell)}} |a_k/d_k\rangle_{A_i/D_i} |L_k\rangle_{c_i} |w, \ell \bmod 2\rangle_{B_k} + \sum_k^{r_{(w,\ell)}} |a_k/d_k\rangle_{A_i/D_i} |R_k\rangle_{c_i} |w, \ell \bmod 2 + 1\rangle_{B_k} \right) ; \forall A_i/D_i
	\end{align*}
	
	The entire retrieving operation of $\tilde{\mathcal{R}}_{(w+1|w)}^{i\ \dagger}$ from the daughter nodes $(w + 1, \ell \bmod 2)$ and $(w + 1, \ell \bmod 2 + 1)$ to the parent node $(w, \ell)$ for the $i^{\text{th}}$ walker can be summarized as:
	
	\begin{align*}
		&\frac{1}{\sqrt{\aleph}} \left( \sum_k^{l_{(w+1,\ell)}} |a_k/d_k\rangle_{A_i/D_i} |L_k\rangle_{c_i} |w + 1, \ell \bmod 2 + 1\rangle_{B_k} + \sum_k^{r_{(w+1,\ell)}} (-1)^{l_{(w,\ell)}} \times |a_k/d_k\rangle_{A_i/D_i} |R_k\rangle_{c_i} |w + 1, \ell \bmod 2\rangle_{B_k} \right) \\
		&\xrightarrow{\tilde{\mathcal{R}}_{(w|w+1)}^{i\ \dagger}} \frac{1}{\sqrt{\aleph}} \left( \sum_k^{l_{(w,\ell)}} |a_k/d_k\rangle_{A_i/D_i} |L_k\rangle_{c_i} |w, \ell \bmod 2\rangle_{B_k} + \sum_k^{r_{(w,\ell)}} |a_k/d_k\rangle_{A_i/D_i} |R_k\rangle_{c_i} |w, \ell \bmod 2 + 1\rangle_{B_k} \right) ; \forall A_i/D_i ; \\
		&\tilde{\mathcal{R}}_{(w|w+1)}^{i\ \dagger} = \sum_{\ell=0}^{2^w-1} \hat{\Psi}_{(w,\ell)}^\dagger \mathcal{H}_{(w,\ell)}^\dagger
	\end{align*}
	
	Therefore, the entire retrieving operation $\tilde{\mathcal{R}}^\dagger$ of $n + m$ walkers from node $(n, k)$ ($2^n$ memory addresses) back to the root node $(0,0)$ is given by applying the tensor product of $\tilde{\mathcal{R}}_{(w|w+1)}^{i\ \dagger}$ recursively for each level of the binary tree to all $n + m$ walkers as follows:
	
	\begin{align*}
		&\sum_k^j (|a_k\rangle_{A_{n-1}} \otimes .. \otimes |a_k\rangle_{A_0})(|d_k\rangle_{D_{m-1}} \otimes .. \otimes |d_k\rangle_{D_0})(|c_k\rangle_{c_{n+m-1}} \otimes \dots \otimes |c_k\rangle_{c_0})|n, k\rangle_{B_k} \\
		&\xrightarrow{\tilde{\mathcal{R}}^\dagger} \frac{1}{\sqrt{\aleph}} \sum_k (|a_k\rangle_{A_{n-1}} \otimes .. \otimes |a_k\rangle_{A_0})(|d_k\rangle_{D_{m-1}} \otimes .. \otimes |d_k\rangle_{D_0})(|L_k\rangle_{c_{n+m-1}} \otimes \dots \otimes |L_k\rangle_{c_0})|0,0\rangle_{B_j} \\
		&; \ \tilde{\mathcal{R}}^\dagger = \bigotimes_{i=0}^{n+m-1} \tilde{\mathcal{R}}_{(0|1)}^{i\ \dagger} \tilde{\mathcal{R}}_{(1|2)}^{i\ \dagger} \dots \tilde{\mathcal{R}}_{(n-1|n)}^{i\ \dagger}
	\end{align*}
	
	We now apply these relations to our qRAM example across six layers of three-level binary trees, retrieving a superposition of three-qubit address state walkers, $\frac{1}{\sqrt{3}}(|0\rangle_{A_0} \otimes |1\rangle_{A_1} \otimes |0\rangle_{A_2} + |1\rangle_{A_0} \otimes |0\rangle_{A_1} \otimes |0\rangle_{A_2} + |1\rangle_{A_0} \otimes |1\rangle_{A_1} \otimes |1\rangle_{A_2})$, and three-qubit data state walkers, $\frac{1}{\sqrt{3}}(|0\rangle_{D_0} \otimes |1\rangle_{D_1} \otimes |1\rangle_{D_2} + |1\rangle_{D_0} \otimes |0\rangle_{D_1} \otimes |1\rangle_{D_2} + |1\rangle_{D_0} \otimes |1\rangle_{D_1} \otimes |0\rangle_{D_2})$, back to the root node $(0,0)$. (See Fig.~\ref{fig:fig6}) For the superposition state of all six walkers coming from the memory addresses, the total retrieval is as follows:
	
	\begin{align*}
		&\frac{1}{\sqrt{3}} \Big( (|0\rangle_{A_0} \otimes |1\rangle_{A_1} \otimes |0\rangle_{A_2})(|0\rangle_{D_0} \otimes |1\rangle_{D_1} \otimes |1\rangle_{D_2})(|L\rangle_{c_0} \otimes .. \otimes |L\rangle_{c_5})|3,2\rangle_{B_0} \Big) \\
		&+ \Big( (|1\rangle_{A_0} \otimes |0\rangle_{A_1} \otimes |0\rangle_{A_2})(|1\rangle_{D_0} \otimes |0\rangle_{D_1} \otimes |1\rangle_{D_2})(|L\rangle_{c_0} \otimes .. \otimes |L\rangle_{c_5})|3,4\rangle_{B_1} \Big) \\
		&+ \Big( (|1\rangle_{A_0} \otimes |1\rangle_{A_1} \otimes |1\rangle_{A_2})(|1\rangle_{D_0} \otimes |1\rangle_{D_1} \otimes |0\rangle_{D_2})(|R\rangle_{c_0} \otimes .. \otimes |R\rangle_{c_5})|3,7\rangle_{B_2} \Big) \\
		&\xrightarrow{\bigotimes_{i=0}^5 \tilde{\mathcal{R}}_{(0|1)}^{i\ \dagger} \tilde{\mathcal{R}}_{(1|2)}^{i\ \dagger} \tilde{\mathcal{R}}_{(2|3)}^{i\ \dagger}} \\
		&\frac{1}{\sqrt{3}} \Big( \big( (|0\rangle_{A_0} \otimes |1\rangle_{A_1} \otimes |0\rangle_{A_2})(|0\rangle_{D_0} \otimes |1\rangle_{D_1} \otimes |1\rangle_{D_2})(|L\rangle_{c_0} \otimes .. \otimes |L\rangle_{c_5}) \big) \\
		&+ \big( (|1\rangle_{A_0} \otimes |0\rangle_{A_1} \otimes |0\rangle_{A_2})(|1\rangle_{D_0} \otimes |0\rangle_{D_1} \otimes |1\rangle_{D_2})(|L\rangle_{c_0} \otimes .. \otimes |L\rangle_{c_5}) \big) \\
		&+ \big( (|1\rangle_{A_0} \otimes |1\rangle_{A_1} \otimes |1\rangle_{A_2})(|1\rangle_{D_0} \otimes |1\rangle_{D_1} \otimes |0\rangle_{D_2})(|L\rangle_{c_0} \otimes .. \otimes |L\rangle_{c_5}) \big) \Big) |0,0\rangle_{B_0}
	\end{align*}
	
	\section{Results and discussions}
	
	\subsection{Experimental Considerations for Routing Operator}
	
	\begin{figure}[htbp]
		\centering
		\includegraphics[width=0.8\linewidth]{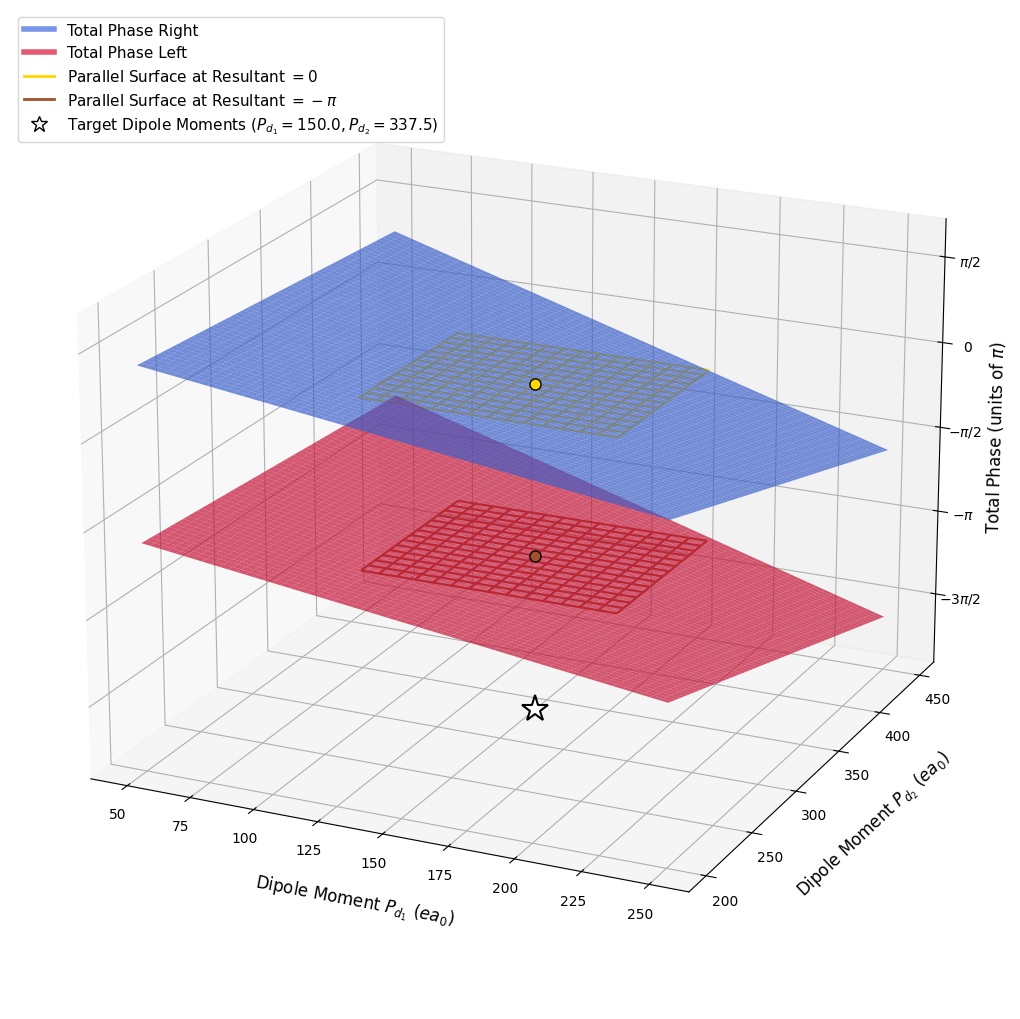}
		\caption{This graph plots polarization-dependent phase generation, displaying total phase acquired by right circularly polarized probes on a blue surface and left probes on a red surface, as a function of dipole moments $P_{d_1}$ and $P_{d_2}$. Calibration sets linear phase ($\pm i\Delta L$) to $\pm \pi/2$ and nonlinear polariton interaction-induced phase $c_j/\hbar\omega^2 v$ to $\pi/2$. Starred markers indicate target dipole moments $P_{d_1}=150ea_0(n=10,l=0)$ and $P_{d_2}=337.5ea_0(n=10,l=0 \text{ and } n=15,l=0)$ satisfying EIT constraints.}
		\label{fig:fig7}
	\end{figure}
	
	Experimentally, the routing operator steers oppositely polarized signal and probe pulses, returning the probe with a relative nonlinear phase shift. We set the output linear phase $\pm i \Delta L$ for the left (right) circularly polarized probes to $\pm \frac{\pi}{2}$. This is achieved using a magnetic field $\sim 10^{-9} T$, gyromagnetic ratio $g_F=1 / 2$, magnetic quantum number $M_F= \pm 1$, and waveguide length $L=1 c m$, assuming detunings $\delta_{1,2}=\omega_p-\omega_{e_{1,2} g_{1,2}}-k_p v \mp \Delta$ approach resonance ( $\omega_p \rightarrow \omega_{e_{1,2} g_{1,2}}$ ). Concurrently, we set the polariton interaction-induced nonlinear phase to $\frac{C_{i j}}{\hbar w^2 v}=\frac{\pi}{2}$. This dipole-dipole phase shift must satisfy three conditions \cite{shahmoon2011, petrosyan2012, friedler2005_2, lukin2003}: (i) The probe pulse duration $T$ must exceed the inverse EIT bandwidth, $T>\left(\gamma_{g e_l} \sqrt{\kappa_l L}\right) /\left|\Omega_l\right|^2$, where $\gamma_{g e_l}$ is the relaxation rate and $\kappa_l L=\frac{3 \lambda_l^2}{2 \pi} \frac{N}{\left[\pi\left(w_a^2+w_l^2\right)\right]}$ is the absorption coefficient. Satisfying this requires $t_{\text {out }} \gtrsim v_l / L$, with group velocity $v_l= 2\left|\Omega_l\right|^2 /\left(\kappa_l \gamma_{\text {gel }}\right)$ \cite{lukin2003, zhao2022}. (ii) The total dipole-dipole phase shift must remain below the EIT bandwidth, $\frac{C_{i j}}{\hbar w^2 v}< \left|\Omega_l\right|^2 /\left(\gamma_{g e_l} \sqrt{\kappa_l L}\right)$, which, combined with condition (i), bounds the shift to $\frac{C_{i j}}{\hbar w^2 v} \leq \frac{t_{\text {out }}\left|\Omega_l\right|^2}{\left(\gamma_{g e_l} \sqrt{\kappa_l L}\right)} \lesssim \frac{1}{2} \sqrt{\kappa_l L}$. (iii) The pulse propagation time $t_{\text {out }}$ inside the waveguide is constrained by the relaxation rate $\gamma_{g d_l}$ \cite{petrosyan2012}. To realize $\frac{C_{i j}}{\hbar w^2 v} \approx \frac{\pi}{2}$, we assume a cavity atom count $N \sim 10^4$, optical depth $\kappa_l L \sim 600$, control laser Rabi frequencies $\Omega_l \sim 10^6 \mathrm{rad} / \mathrm{s}$ (yielding $v_l \sim 100 \mathrm{~m} / \mathrm{s}$ ), pulse bandwidth $T^{-1} \sim 10^4$, waveguide length $L \sim 1 \mathrm{~cm}$, transverse mode Gaussian width $w_f \sim 2 \mu m$, and atomic cavity width $w_a \sim 2 \mu m$. We select Rydberg states with dipole moments $P_{d_1}=150 e a_0(n=10, l=0)$ and $P_{d_2}=337.5 e a_0(n=10, l=0 \text{ and } n=15, l=0)$ for the address and signal walker probes, respectively ($a_0$ is the Bohr radius). (See Fig.~\ref{fig:fig7})
	\FloatBarrier
	\subsection{Scalability and complexity of the QRAM Architecture}
	\begin{figure}[htbp]
		\centering
		\includegraphics[width=\textwidth]{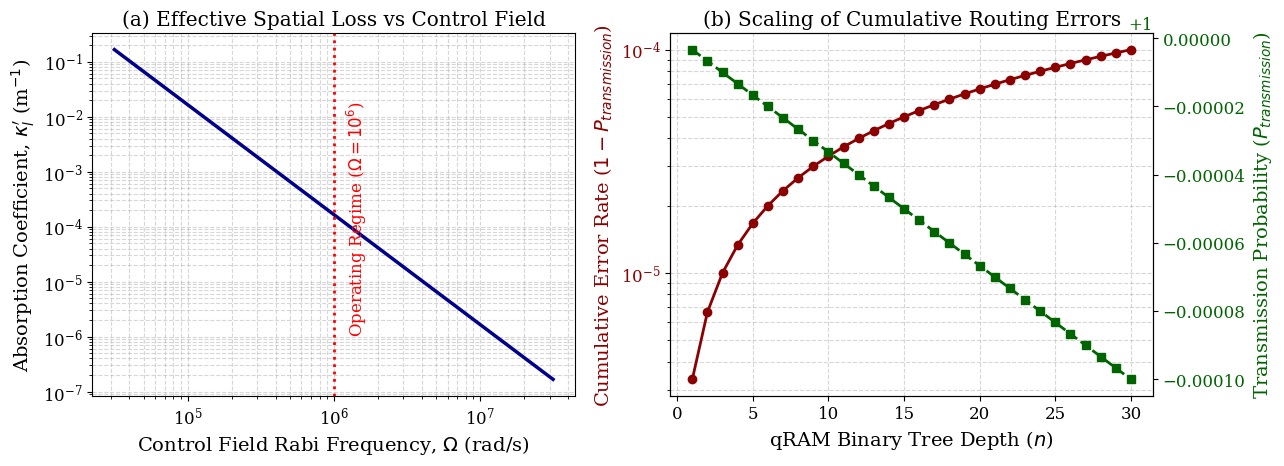}
		\caption{Scaling viability of the qRAM architecture. (a) plots the effective spatial linear absorption coefficient versus control field Rabi frequency, demonstrating attenuation suppression at the targeted operating regime. (b) plots the cumulative routing error rate and transmission probability across binary tree depths, confirming successful preservation of encoded phase information across highly scaled memory addresses.}
		\label{fig:fig8}
	\end{figure}
	
	The transmission probability per node is $P_{\text{transmission}} = |\Psi_l|^2 \exp(-2\kappa_l' z)$. Traversing $2n$ total nodes for depth $n$ yields $P_{\text{transmission}} \propto \exp(-2n\kappa_l' L)$, where node length $L=0.01$ m. Under strict EIT resonance, the effective spatial absorption coefficient is $\kappa_l' \approx \frac{g^2 N^2}{c|\Omega|^2} \gamma_c$. Using experimentally viable parameters ($\Omega=10^6$ rad/s, $g=10^5$ rad/s, $N=10^4$, $\gamma_c=10^3$ s$^{-1}$), this coefficient is heavily suppressed to $\kappa_l' \approx 1.67 \times 10^{-4}$ m$^{-1}$. (See Fig.~\ref{fig:fig8}(a)) For a highly scaled $n=20$ tree (capable of addressing over one million memory cells), cumulative optical loss is $2n\kappa'L \approx 6.67 \times 10^{-5}$, yielding probability $P_{\text{transmission}} \approx 99.993\%$. The routing error rate remains below $10^{-6}$ for depth $n=30$. (See Fig.~\ref{fig:fig8}(b)) Therefore, strong control regime massively suppresses the spatial loss of the probe and successfully preserves dual-rail phase information across $\mathcal{O}(n)$ operations without intermediate error correction. Note this is only the spatial loss faced by the walker after transmitted from the routing nodes consisting of Rydberg atomic cell; the walker is independently subjected to the waveguide loss.
	
	As previously stated, earlier schemes suffer from physical bottlenecks: fanout routing needs exponential gates, whereas Giovannetti's bucket brigade demands error-prone active qutrits \cite{giovannetti2008_1, giovannetti2008_2, giovannetti2008_3}. In contrast, our architecture completely optimizes scaling by mapping Asaka's discrete-time quantum walk onto a physically realizable EIT Rydberg ensemble \cite{asaka2021, asaka2023_1, asaka2023_2}. Spatially, a bus of exactly $n+m$ quantum walkers traverses the binary tree. To route $n+m$ walkers coherently to identical daughter nodes, positional information must be encoded into polarization states $|L\rangle$ or $|R\rangle$ before encountering the operator $\Psi(w,l)$ as per the address bit $a_{n-1-l}$ of the corresponding level $l$. Sequentially imparting address bit $a_{n-1-l}$ through paired interactions creates a linear $\mathcal{O}(n+m)$ bottleneck. Instead, parallel modified Hadamard gates $H(w,l)$ encode the $n+m$ walkers approaching $2^p$ locations. This saturates the bus in strictly $p=\log_2(n+m)$ steps, bounding the complexity of routing information across all walkers at any single node to exactly $\mathcal{O}(\log(n+m))$. Traversing $n$ tree levels requires executing this cascade $n$ times, yielding $\mathcal{O}(n\log(n+m))$ routing complexity. Subsequently, the querying phase transfers stored data to $m$ walkers using simultaneous dual-rail CNOT operations across multiple rails in strictly $\mathcal{O}(1)$ time. Finally, symmetric retrieval requires an additional $\mathcal{O}(n\log(n+m))$ steps. Summing these three continuous operational phases yields a total temporal complexity of precisely $\mathcal{O}(n\log(n+m))$. Therefore, it is concluded that replacing active nodes is extremely scalable and thoroughly successful in execution.
	\FloatBarrier
	
	\section{Conclusion}
	
	We propose a quantum random access memory (qRAM) architecture utilizing quantum random walks (QRWs) to mitigate exponential gate overhead and decoherence. By leveraging strong photon-photon interactions within an EIT-based Rydberg atomic ensemble inside hollow-core waveguides, we replace active qutrit nodes with phase-encoded quantum walkers. A longitudinal magnetic field and strong Rydberg dipole-dipole interactions create a routing operator that imparts precise phase shifts, steering circularly polarized probes into phase-selective waveguides. This framework executes three core operations: routing walkers via modified Hadamard gates, querying $m$-qubit data via dual-rail CNOT operations, and retrieving walkers using conjugate operators, restricting operational steps to $O(n \log (n+m))$, from $2^n$ locations using an $n$-qubit address, requiring only $O(n+m)$ walkers and $O\left(2^n(n+m)\right)$ gates. Ultimately, this scalable QRW approach establishes a practical pathway for advancing exponentially faster quantum algorithms in artificial intelligence and machine learning.

	\section*{Data Availability}
	Not Applicable.
	
	\section*{Conflict of Interest}
	The authors have no relevant financial or non-financial interests to disclose.
	
	\section*{Authors’ Contribution}
	All authors contributed equally.
	
	\bibliographystyle{apsrev4-2} 
	\bibliography{Quantum_Random_Access_Memory_Implementation_Using_Photon_Photon_Interaction_in_Rydberg_Atomic_Ensemble}

\end{document}